\documentclass[12pt,a4paper]{article}

\usepackage{amsmath,amssymb,slashed,cancel}
\usepackage{cite,tabularx}
\usepackage{subcaption}
\usepackage{graphicx,color}
\usepackage[normalem]{ulem}
\usepackage{mathtools}
\usepackage{enumerate}
\usepackage{ascmac}
\usepackage{enumitem}
\usepackage{float}
\usepackage{placeins}
\allowdisplaybreaks

\usepackage[height=8.85in,width=6.4in]{geometry}
\renewcommand{\baselinestretch}{1.1}
\usepackage{comment}
\newcommand\package[2][\relax]{\texttt{#2\ifx#1\relax\relax\relax\else\,\linebreak[0]#1\fi}}

\numberwithin{equation}{section} % Eq.(Sec.eq.)
 
\def\beq#1\eeq{\begin{align}#1\end{align}}

\definecolor{BlueViolet}{rgb}{0.2, 0.00, 0.7}
\definecolor{Blue}{rgb}{0.15, 0.00, 0.9}
\usepackage[colorlinks=true,linkcolor=Blue,citecolor=Blue,urlcolor=BlueViolet]{hyperref}

\begin{document}
\begin{titlepage}
\setcounter{page}{0} % to set page=0 of the title-page

\begin{center}

\vskip .55in

\begingroup
\centering
\large\bf RGE effects on new physics searches via gravitational waves
\endgroup

\vskip .4in

{
Katsuya Hashino$^{\rm (a)}$ and Daiki Ueda$^{\rm (b)}$
}

\vskip 0.4in

\begingroup\small
\begin{minipage}[t]{0.9\textwidth}
\centering\renewcommand{\arraystretch}{0.9}
\begin{tabular}{c@{\,}l}
$^{\rm(a)}$
& National Institute of Technology, Fukushima College, Nagao 30\\
&Taira-Kamiarakawa, Iwaki, Fukushima 970–8034, Japan
\\[2mm]
$^{\rm(b)}$
& Physics Department, Technion \text{--} Israel Institute of Technology
\\
&Technion city, Haifa 3200003, Israel\\
\end{tabular}
\end{minipage}
\endgroup

\end{center}

\vskip .4in

\begin{abstract}\noindent
Gravitational wave (GW) observations offer a promising probe of new physics associated with a first-order electroweak phase transition. 
Precision studies of the Higgs potential, including Fisher matrix analyses, have been extensively conducted in this context.
However, significant theoretical uncertainties in the GW spectrum, particularly those due to renormalization scale dependence in the conventional daisy-resummed approach, have cast doubt on the reliability of such precision measurements. 
These uncertainties have been highlighted using the Standard Model Effective Field Theory (SMEFT) as a benchmark.
To address these issues, we revisit Fisher matrix analyses based on the daisy-resummed approach, explicitly incorporating renormalization scale uncertainties. 
We then reassess the prospects for precise new physics measurements using GW observations.
Adopting the SMEFT as a benchmark, we study the effects of one-loop RGE running of dimension-six operators on the Higgs effective potential via the Higgs self-couplings, top Yukawa coupling, and gauge couplings, in addition to the SMEFT tree-level effects. 
We find that future GW observations can remain sensitive to various dimension-six SMEFT effects, even in the presence of renormalization scale uncertainties, provided that the SMEFT $(H^{\dagger}H)^3$ operator is precisely measured, {\it e.g.}, by future collider experiments.
\end{abstract}
\end{titlepage}

\setcounter{page}{1}
\renewcommand{\thefootnote}{\#\arabic{footnote}}
\setcounter{footnote}{0}

%%%%%%%%%%%%%%%%%%%%%%%%%%%%%%%%%%%%%%%%%%%%%%%%%%%%%%%%
\begingroup
\renewcommand{\baselinestretch}{1} % this command affects only in this "\begingroup-\endgroup"
\setlength{\parskip}{2pt}          % this command affects only in this "\begingroup-\endgroup"
\hrule
\tableofcontents
\vskip .2in
\hrule
\vskip .4in
\endgroup
%%%%%%%%%%%%%%%%%%%%%%%%%%%%%%%%%%%%%%%%%%%%%%%%%%%%%%%%

\section{Introduction}\label{sec:intro}

%(1) NP effects on Higgs potential
The discovery of the Higgs boson at the CERN Large Hadron Collider (LHC)~\cite{ATLAS:2012yve,CMS:2012qbp}, along with precise measurements of its properties consistent with Standard Model (SM) predictions, has significantly reinforced the SM framework.
Despite this success, the exact form of the Higgs potential remains elusive.
Unraveling the nature of the electroweak phase transition is, therefore, a central open question in particle physics.
In particular, a strongly first-order electroweak phase transition (SFO-EWPT) can be a key ingredient for realizing electroweak baryogenesis, which accounts for the observed baryon asymmetry of the universe~\cite{Gavela:1993ts,Konstandin:2003dx,Kuzmin:1985mm}.
However, in the SM, such a transition only occurs if the Higgs mass is below approximately 65 GeV~\cite{Kajantie:1995kf,Kajantie:1996mn,Kajantie:1996qd,Csikor:1998eu,DOnofrio:2015gop}, far below the measured value of 125 GeV~\cite{ParticleDataGroup:2024cfk}.
This mismatch has sparked increasing interest in how new physics (NP) might modify the Higgs potential.
Despite this, the continued absence of direct evidence for new particles at the LHC suggests that the NP scale may lie well above the electroweak symmetry breaking scale.

In light of this situation, the EFT framework offers a natural approach to encode the effects of NP via higher-dimensional operators.
The SMEFT~\cite{Grzadkowski:2010es,Jenkins:2013zja,Jenkins:2013wua,Alonso:2013hga} provides a well-established setting in which such effects are systematically captured through operators involving SM fields.
Among them, the dimension-six $(H^{\dagger}H)^3$ operator has been extensively studied~\cite{Ellis:2018mja,Grojean:2004xa,Zhang:1992fs,Bodeker:2004ws,Binetruy:2012ze,Delaunay:2007wb,Huber:2013kj,Konstandin:2014zta,Damgaard:2015con,Harman:2015gif,Balazs:2016yvi,deVries:2017ncy,Cai:2017tmh,Chala:2018ari,Dorsch:2018pat,DeVries:2018aul,Chala:2019rfk,Ellis:2019flb,Zhou:2019uzq,Kanemura:2020yyr,Phong:2020ybr,Wang:2020zlf,Wang:2020jrd,Kanemura:2021fvp,Lewicki:2021pgr,Hashino:2021qoq,Kanemura:2022txx,Anisha:2022hgv,Huang:2015izx,Cao:2017oez,Huang:2016odd,Croon:2020cgk,Hashino:2022ghd,Banerjee:2024qiu,Gazi:2024boc,Camargo-Molina:2024sde,Bernardo:2025vkz,Chala:2025aiz,Zhu:2025pht} in relation to achieving a SFO-EWPT.
Additional dimension-six SMEFT operators have also been investigated in the context of electroweak baryogenesis~\cite{Bodeker:2004ws,Balazs:2016yvi,Huang:2015izx,Cao:2017oez,Huang:2016odd,Huber:2006ri}.
On the experimental side, growing efforts aim to constrain SMEFT Wilson coefficients using data from both existing and future collider experiments.
These include high-luminosity LHC~\cite{Cepeda:2019klc}, International Linear Collider (ILC)~\cite{LCCPhysicsWorkingGroup:2019fvj,ILCInternationalDevelopmentTeam:2022izu}, Compact LInear Collider (CLIC)~\cite{CLIC:2018fvx}, Future Circular Collider of electrons and positrons (FCC-ee)~\cite{FCC:2018byv}, and Circular
Electron Positron Collider (CEPC)~\cite{An:2018dwb}.
Moreover, the SFO-EWPT, {\it e.g.}, via the SMEFT, is expected to generate a stochastic gravitational wave (GW) background with a peak in the milli- to deci-Hertz range, accessible to future space-based GW detectors such as LISA~\cite{LISA:2017pwj}, DECIGO~\cite{Seto:2005qy}, and BBO~\cite{BBO}.

However, the SFO-EWPT via the SMEFT $(H^{\dagger}H)^3$ operator has known limitations regarding the validity range of the EFT description.
In Ref.~\cite{Damgaard:2015con}, the validity of the SMEFT is examined by comparing the SFO-EWPT in the SM extended with a singlet scalar boson to that with a $(H^{\dagger}H)^3$ operator.
It is shown that the applicability of the SMEFT is restricted to a small region of the full singlet model parameter space.
Furthermore, in Ref.~\cite{Postma:2020toi}, the validity of the SMEFT is analyzed in a nearly model-independent manner based on the covariant derivative expansion.
It is demonstrated that, in a wide range of theories, the SFO-EWPT cannot be accurately captured by the SMEFT truncated at dimension-six operators.
An exception to this general argument is the singlet-extended SM.

Although the reliability of the SFO-EWPT description via the SMEFT $(H^{\dagger}H)^3$ operator has been examined, the EWPT in the SMEFT exhibits qualitative features similar to those found in single-step transitions of scalar-extended SM, such as the two-Higgs-doublet model~\cite{Caprini:2019egz}.
Furthermore, the dimension-six SMEFT operator $(H^{\dagger}H)^3$ captures the effects of new physics on the Higgs potential in a model-independent manner.
Consequently, the SMEFT remains a valuable benchmark for assessing theoretical uncertainties~\cite{Croon:2020cgk}\footnote{Also see Ref.~\cite{Athron:2022jyi} for the theoretical uncertainties of the SM extended by a real scalar singlet with $Z_2$ symmetry.
Relatedly, uncertainties and potential issues in treating the gravitational wave spectrum under supercooling have been studied in Refs.~\cite{Athron:2023rfq,Athron:2022mmm}.
}.
In particular, it highlights the sensitivity to the renormalization scale in the context of GW sources.
Ref.~\cite{Croon:2020cgk} demonstrates that the conventional daisy-resummed approach exhibits significant uncertainties associated with renormalization scale dependence, using the SMEFT as a diagnostic tool.
Furthermore, it is pointed out that both the amplitude and peak frequency of the resulting GW spectrum can only be predicted with an accuracy that strongly depends on the computational method employed, such as the daisy-resummed approach.
As a result, the reliability of new physics predictions via GW observations based on the daisy-resummed method was questioned, again using the SMEFT as a reference framework.

Relatedly, earlier studies, including Ref.~\cite{Croon:2020cgk}, have primarily focused on identifying regions of parameter space that could lead to a detectable GW signal.
More recently, Ref.~\cite{Hashino:2022ghd} explored how precisely the effects of SMEFT operators could be measured in the event of a GW detection.
In that study, three types of effects from dimension-six SMEFT operators on the GW spectrum, induced via a SFO-EWPT, were investigated: (i) the tree-level contribution from the $(H^{\dagger}H)^3$ operator, (ii) the tree-level contribution from the wave function renormalization of the Higgs field, and (iii) the one-loop contribution from the top quark.
The sensitivity of future GW experiments to these effects was quantitatively assessed using a Fisher matrix analysis.
It was found that, provided that the SFO-EWPT is driven by (i) the operator $(H^{\dagger}H)^3$, future GW observations could exhibit significant sensitivity to additional effects from (ii) wave function renormalization and (iii) top-quark loops.
These studies suggest that GW observations have the potential to precisely probe new physics.
However, it has been pointed out that the commonly used daisy-resummed approach suffers from significant theoretical uncertainties, particularly due to renormalization scale dependence, which can lead to variations of a few orders of magnitude in the predicted peak GW amplitude.
Therefore, at this stage, an open question is whether future GW observations can maintain their precision in probing new physics effects, such as those arising from SMEFT operators, even when renormalization scale dependence is adequately considered in the daisy-resummed approach.

In this work, we address this open question and demonstrate that future GW observations can remain highly sensitive to various dimension-six SMEFT operators, provided that other experiments, including future collider~\cite{DeBlas:2019qco,deBlas:2019rxi}, precisely determine the effects of the $(H^{\dagger}H)^3$ operator.
We adopt the SMEFT as a benchmark framework and comprehensively investigate the dimension-six operators that can affect the Higgs effective potential through one-loop RGE corrections involving the Higgs self-couplings, the top Yukawa coupling, and the gauge couplings.
We focus on a scenario in which the $(H^{\dagger}H)^3$ operator drives the SFO-EWPT, while other dimension-six SMEFT operators introduce small shifts in the Higgs potential.
We evaluate the impact of these SMEFT effects on the GW spectrum and perform a Fisher matrix analysis at two different renormalization scales, $\overline{\mu}_{\rm PT}=2\pi T$ and $T/2$, to clarify the expected sensitivity of future GW observations and their dependence on the renormalization scale.
Furthermore, we repeat the analysis while varying the bubble wall velocity $v_b$ to examine the associated theoretical uncertainties.

This paper is organized as follows:
In Section~\ref{sec:form}, we present the RGE effects of dimension-six SMEFT operators on the Higgs potential at the one-loop level.
We also provide the one-loop Coleman-Weinberg potential incorporating the dimension-six SMEFT operators, along with the thermal effective potential obtained using the daisy-resummed approach.
In particular, subsection~\ref{sec:proc} summarizes the procedures for analyzing the SFO-EWPT within the daisy-resummed framework; see also Fig.~\ref{fig:sch} for further details.
In Section~\ref{sec:GW_sec}, we analyze the effects of the SMEFT dimension-six operators on the GW spectrum from the SFO-EWPT.
Using a Fisher matrix analysis, we investigate how precisely future GW observations can probe the SMEFT dimension-six operators, while taking into account the renormalization scale uncertainties.
See Figs.~\ref{fig:contour}, \ref{fig:DECIGO05}, \ref{fig:DECIGO02}, and \ref{fig:DECIGO1} for the main results.
Finally, we conclude with a summary of this work in Section~\ref{sec:summary}.

\section{Higgs effective potential}\label{sec:form}
We first examine the SMEFT dimension-six operator effects, especially on the Higgs potential.
In subsection~\ref{sec:matching_tree}, relations between parameters in the Lagrangian and physical observables are provided in the SMEFT at the tree level. 
These relations are extended up to the one-loop level in subsection~\ref{sec:matching_loop}, and relations between $\overline{{\rm MS}}$-parameters and physical observables are provided at the one-loop level.
While subsection~\ref{sec:CW} lists the one-loop Coleman-Weinberg potential, including the SMEFT dimension-six operator effects. 
After presenting the one-loop thermal effective Higgs potential in subsection~\ref{sec:thermal}, we summarize in subsection~\ref{sec:proc} the procedures based on daisy resummation used in this work to evaluate the EWPT.

\subsection{Preliminary}
\label{sec:matching_tree}
To analyze the effects of RG scale dependence on the search for NP via GW observations, we follow Ref.~\cite{Croon:2020cgk} and adopt the following SMEFT Lagrangian as a benchmark model:
\begin{align}
    \mathcal{L}_{\rm SMEFT} = \mathcal{L}_{\rm SM} +\sum_i C_i \mathcal{O}_i, \label{eq:full_L}
\end{align}
where the first term corresponds to the SM Lagrangian, while the second term represents higher-dimensional operators composed of the SM fields.
In this paper, we incorporate SMEFT effects by including dimension-six operators and their one-loop RGE corrections.
The operators considered in the context of the SFO-EWPT are summarized in Table~\ref{tab:dim6}.
\begin{table}
\begin{center}
\small
\begin{minipage}[t]{5.0cm}
\renewcommand{\arraystretch}{1.5}
\begin{tabular}[t]{c|c}
\multicolumn{2}{c}{$X^3$} \\
\hline\hline
$\mathcal{O}_W$                & $\epsilon^{IJK} W_\mu^{I\nu} W_\nu^{J\rho} W_\rho^{K\mu}$ \\
\end{tabular}
\end{minipage}
\begin{minipage}[t]{5.6cm}
\renewcommand{\arraystretch}{1.5}
\begin{tabular}[t]{c|c}
\multicolumn{2}{c}{$ H^6~{\rm and}~H^4 D^2$} \\
\hline\hline
$\mathcal{O}_H$ & $(H^\dag H)^3$\\
$\mathcal{O}_{H\Box}$ & $(H^\dag H)\Box(H^\dag H)$ \\
$\mathcal{O}_{H D}$   & $\ \left(H^\dag D_\mu H\right)^* \left(H^\dag D_\mu H\right)$
\end{tabular}
\end{minipage}
\begin{minipage}[t]{2.7cm}

\renewcommand{\arraystretch}{1.5}
\begin{tabular}[t]{c|c}
\multicolumn{2}{c}{$\psi^2H^3 + \hbox{h.c.}$} \\
\hline\hline
$\mathcal{O}_{uH}$          & $(H^\dag H)(\bar q_p u_r \widetilde H )$ \\
\end{tabular}
\end{minipage}

\vspace{0.25cm}

\begin{minipage}[t]{5.0cm}
\renewcommand{\arraystretch}{1.5}
\begin{tabular}[t]{c|c}
\multicolumn{2}{c}{$X^2H^2$} \\
\hline\hline
$\mathcal{O}_{H G}$     & $H^\dag H\, G^A_{\mu\nu} G^{A\mu\nu}$ \\
$\mathcal{O}_{H\widetilde G}$         & $H^\dag H\, \widetilde G^A_{\mu\nu} G^{A\mu\nu}$ \\
$\mathcal{O}_{H W}$     & $H^\dag H\, W^I_{\mu\nu} W^{I\mu\nu}$ \\
$\mathcal{O}_{H\widetilde W}$         & $H^\dag H\, \widetilde W^I_{\mu\nu} W^{I\mu\nu}$ \\
$\mathcal{O}_{H B}$     & $ H^\dag H\, B_{\mu\nu} B^{\mu\nu}$ \\
$\mathcal{O}_{H\widetilde B}$         & $H^\dag H\, \widetilde B_{\mu\nu} B^{\mu\nu}$ \\
$\mathcal{O}_{H WB}$     & $ H^\dag \tau^I H\, W^I_{\mu\nu} B^{\mu\nu}$ \\
$\mathcal{O}_{H \widetilde W B}$         & $H^\dag \tau^I H\, \widetilde W^I_{\mu\nu} B^{\mu\nu}$
\end{tabular}
\end{minipage}
\begin{minipage}[t]{5.2cm}
\renewcommand{\arraystretch}{1.5}
\begin{tabular}[t]{c|c}
\multicolumn{2}{c}{$\psi^2 XH+\hbox{h.c.}$} \\
\hline\hline
$\mathcal{O}_{uG}$        & $(\bar q_p \sigma^{\mu\nu} T^A u_r) \widetilde H \, G_{\mu\nu}^A$ \\
$\mathcal{O}_{uW}$        & $(\bar q_p \sigma^{\mu\nu} u_r) \tau^I \widetilde H \, W_{\mu\nu}^I$ \\
$\mathcal{O}_{uB}$        & $(\bar q_p \sigma^{\mu\nu} u_r) \widetilde H \, B_{\mu\nu}$ 
\end{tabular}
\end{minipage}
\begin{minipage}[t]{5.0cm}
\renewcommand{\arraystretch}{1.5}
\begin{tabular}[t]{c|c}
\multicolumn{2}{c}{$\psi^2H^2 D$} \\
\hline\hline
$\mathcal{O}_{H l}^{(3)}$      & $(H^\dag i\overleftrightarrow{D}^I_\mu H)(\bar l_p \tau^I \gamma^\mu l_r)$\\
$\mathcal{O}_{H q}^{(1)}$      & $(H^\dag i\overleftrightarrow{D}_\mu H)(\bar q_p \gamma^\mu q_r)$\\
$\mathcal{O}_{H q}^{(3)}$      & $(H^\dag i\overleftrightarrow{D}^I_\mu H)(\bar q_p \tau^I \gamma^\mu q_r)$\\
$\mathcal{O}_{H u}$            & $(H^\dag i\overleftrightarrow{D}_\mu H)(\bar u_p \gamma^\mu u_r)$\\
\end{tabular}
\end{minipage}
\vspace{0.25cm}
\begin{minipage}[t]{4.2cm}

\renewcommand{\arraystretch}{1.5}
\begin{tabular}[t]{c|c}
\multicolumn{2}{c}{$(\bar LL)(\bar LL)$} \\
\hline\hline
$\mathcal{O}_{ll}$        & $(\bar l_p \gamma_\mu l_r)(\bar l_s \gamma^\mu l_t)$ 
\end{tabular}
\end{minipage}
\begin{minipage}[t]{5.2cm}
\renewcommand{\arraystretch}{1.5}
\begin{tabular}[t]{c|c}
\multicolumn{2}{c}{$(\bar LL)(\bar RR)$} \\
\hline\hline
$\mathcal{O}_{qu}^{(1)}$         & $(\bar q_p \gamma_\mu q_r)(\bar u_s \gamma^\mu u_t)$ \\
$\mathcal{O}_{qu}^{(8)}$         & $(\bar q_p \gamma_\mu T^A q_r)(\bar u_s \gamma^\mu T^A u_t)$ \\
\end{tabular}
\end{minipage}
\end{center}
\caption{\label{tab:dim6}
The dimension-six operators in the SMEFT~\cite{Grzadkowski:2010es} that are potentially relevant to the SFO-EWPT are listed.
These operators affect the Higgs potential through both tree-level contributions and one-loop RGE effects involving the Higgs self-couplings $m^2$ and $\lambda$, the top Yukawa coupling $Y_t$, and the gauge couplings $g_s$, $g$, and $g'$.
}
\end{table}

The presence of dimension-six operators can lead to modifications of the SM parameters.
For more details, see, {\it e.g.}, Ref.~\cite{Alonso:2013hga}.
First, at the tree level, we consider the modifications to the parameters appearing in the Higgs potential.
One-loop effects will be examined in subsection~\ref{sec:matching_loop}.
The relevant terms in the SMEFT Lagrangian~\eqref{eq:full_L}, involving only the Higgs fields, are given as follows:
\begin{align}
    \mathcal{L}_{\rm SMEFT}\supset \mathcal{L}_{\rm Higgs}:&=(D_{\mu}H)^{\dagger}(D^{\mu}H) +m^2 H^{\dagger}H -\frac{\lambda}{2} (H^{\dagger}H)^2\notag
    \\
    &+C_H (H^{\dagger}H)^3+C_{H\Box}(H^{\dagger}H)\Box (H^{\dagger}H)+C_{HD} (H^{\dagger}D^{\mu}H)^{\ast} (H^{\dagger} D_{\mu} H)
    ,\label{eq:SMEFT_Higgs}
\end{align}
where the covariant derivative is defined as $D_{\mu} = \partial_{\mu} + i g_s T^A G^A_{\mu} + i g T^I W^I_{\mu} + i g' Y B_{\mu}$,  
where $\{g_s,\, g,\, g'\}$ and $\{G_{\mu}^A,\, W_{\mu}^I,\, B_{\mu}\}$ are the gauge couplings and gauge fields of the $SU(3)_c$, $SU(2)_L$, and $U(1)_Y$ gauge groups, respectively, and the corresponding generators are denoted by $T^A$ and $T^I$. 
The hypercharge assignments for the matter fields are denoted by $Y$.
In the unitary gauge, the Higgs field can be written as $H=\left(0,[1+c_{H,{\rm kin}}]h+v\right)^T/\sqrt{2}$ with $c_{H,{\rm kin}}:= \left(C_{H\Box}-C_{HD}/4\right)v^2$.
The coefficient $c_{H,{\rm kin}}$ is introduced to canonically normalize the kinetic term of the field $h$, and the vacuum expectation value (VEV) $v$ is defined such that the Higgs potential~\eqref{eq:SMEFT_Higgs} is locally minimized at $v$.
Then, the Lagrangian $\mathcal{L}_{\rm Higgs}$ yields
\begin{align}
    \mathcal{L}_{\rm Higgs}\supset \frac{1}{2}(\partial_{\mu}h)^2-V_h,
\end{align}
where terms involving higher derivatives (arising from $C_{H\Box}$ and $C_{HD}$) and gauge fields are omitted for brevity.
The potential $V_h$ is given as,
\begin{align}
    V_h=V+\sum_{i=1}^6a_i h^i,\label{eq:Vh}
\end{align}
with
\begin{align}
    &a_1=-\frac{v}{4} \left[4m^2-2\lambda v^2 +3 C_H v^4 +c_{H,{\rm kin}} (4m^2-2\lambda v^2)\right],
    \\
    &a_2=-\frac{1}{8}\left[4m^2 -6\lambda v^2+15 C_H v^4+2c_{H,{\rm kin}} (4m^2-6\lambda v^2)\right],
    \\
    &a_3=-\frac{v}{2}\left[
    -\lambda+5C_H v^2 -3\lambda c_{H,{\rm kin}}
    \right],
    \\
    &a_4=\frac{1}{8}\left[\lambda-15 C_H v^2+4\lambda c_{H,{\rm kin}}\right],
    \\
    &a_5=-\frac{3v}{4}C_H,
    \\
    &a_6=-\frac{1}{8}C_H.
\end{align}
The first term $V$ in eq.~\eqref{eq:Vh} denotes the zeroth-order term in $h$, defined as
\begin{align}
    V=-\frac{m^2}{2}\phi^2+\frac{\lambda}{8}\phi^4-\frac{C_H}{8}\phi^6,\label{eq:Vtree}
\end{align}
with $\phi=v$.
In what follows, we analyze the EWPT while neglecting terms of second order in the SMEFT Wilson coefficients.
The parameters $m^2$ and $\lambda$ can be determined using the following relations:
\begin{align}
    &\partial_h V_h|_{h=0}=0\Rightarrow 4m^2-2\lambda v^2 +3 C_H v^4 +c_{H,{\rm kin}} (4m^2-2\lambda v^2)=0,\label{eq:dV}
    \\
    &\partial_h^2 V_h|_{h=0}=M_h^2 \Rightarrow -\frac{1}{4}\left(4m^2 -6\lambda v^2+15 C_H v^4+2c_{H,{\rm kin}} (4m^2-6\lambda v^2)\right)=M_h^2,\label{eq:ddV}
\end{align}
where $M_h$ denotes the Higgs boson mass.
To first order in the Wilson coefficients, eqs.~\eqref{eq:dV} and \eqref{eq:ddV} yield
\begin{align}
    m^2&=\frac{1}{2}M_h^2 +\frac{3}{4}C_H v^4-c_{H,{\rm kin}}M_h^2,\label{eq:m2_tree}
    \\
    \lambda&=\frac{M_h^2}{v^2}+3C_H v^2-c_{H,{\rm kin}}\frac{2M_h^2}{v^2},\label{eq:lambda_tree}
\end{align}
where the right-hand sides are expressed in terms of physical observables ($M_h$) and SMEFT Wilson coefficients ($C_H$, $C_{H\Box}$, and $C_{HD}$).
Here, the Higgs VEV $v$ can be determined from $G_F$ (measured in $\mu$ decay, $\mu^- \to e^- + \bar{\nu}_e + \nu_\mu$), taking into account SMEFT effects.  
For convenience, we list in Table~\ref{tab:SMval} the physical observables relevant to our analysis, as given in Ref.~\cite{ParticleDataGroup:2024cfk}.
\begin{table}[htb]
  \begin{center}
    \begin{tabular}{ccccc} \hline\hline 
     $\overline{\rm MS}$-Parameter  & Observables & Central Value 
     \\ \hline
      $m^2$ & $G_F$ & $G_F=1.1664\times 10^{-5}$ GeV$^{-2}$ 
      \\                            
         $\lambda$ & $M_h$ & $M_h=125.20$ GeV
         \\                            
            $g'$ & $M_W$ & $M_W=80.369$ GeV 
            \\  
           $g$ & $M_Z$ & $M_Z=91.188$ GeV 
           \\            
           $g_s$ & Multiple & $\alpha_s(M_Z)=0.1180$ 
           \\          
           $Y_t$ & $M_t$ & $M_t=172.6$ GeV
           \\             
   \hline  \hline  
    \end{tabular}
        \caption{Experimental values for the observables are taken from Ref.~\cite{ParticleDataGroup:2024cfk}. 
    The observables and $\overline{\rm MS}$ parameters are matched at one-loop order at the input scale $\overline{\mu} = M_Z$; see, {\it e.g.}, subsection~\ref{sec:matching_loop}, and Appendix~\ref{sec:MSbar_para}, for the details. 
        }
 \label{tab:SMval}
  \end{center}
\end{table}
Up to the first order of the SMEFT Wilson coefficients, the Higgs VEV can be expressed at the tree-level as
\begin{align}
    v&=v_0 -\frac{1}{4}\left[
    (C_{ll})_{2 112}+ (C_{ll})_{1221}
    -2 \left(
    (C^{(3)}_{Hl})_{11}+ (C^{(3)}_{Hl})_{22}
    \right)
    \right]v_0^3,\label{eq:VEV}
\end{align}
with a physical observable $v_0:=1/\left(2^{1/4}\sqrt{G_F}\right)$, and the SMEFT Wilson coefficients ($(C_{ll})_{2 112}$, $(C_{ll})_{1221}$, $(C^{(3)}_{Hl})_{11}$, and $(C^{(3)}_{Hl})_{22}$).
From these expressions~\eqref{eq:m2_tree}, \eqref{eq:lambda_tree}, and \eqref{eq:VEV}, we find that the Higgs potential can be modified at the tree level by the SMEFT operators.

Second, the SMEFT operators can modify the electroweak gauge boson masses and gauge couplings at the tree level~\cite{Alonso:2013hga}.
Let us focus on the relevant operators to only the gauge fields:
\begin{align}
    \mathcal{L}_{\rm SMEFT}\supset &-\frac{1}{4}G^A_{\mu\nu}G^{A\mu\nu} -\frac{1}{4} W^I_{\mu\nu}W^{I\mu\nu}-\frac{1}{4}B_{\mu\nu}B^{\mu\nu}\notag
    \\
    &+C_{HG} H^{\dagger}H G^A_{\mu\nu} G^{A\mu\nu}+C_{HW} H^{\dagger}H W^I_{\mu\nu}W^{I\mu\nu}\notag
    \\
    &+C_{HB} H^{\dagger}H B_{\mu\nu}B^{\mu\nu}+C_{HWB} H^{\dagger}\tau^I H W^I_{\mu\nu} B^{\mu\nu}.
\end{align}
In the broken phase, the normalized gauge fields are defined as $G^A_{\mu}= \mathcal{G}^A_{\mu}\left(1+C_{HG} v^2\right)$, $W^I_{\mu}=\mathcal{W}^I_{\mu} \left(1+C_{HW} v^2\right)$,  $B_{\mu}=\mathcal{B}_{\mu}\left(1+C_{HB}v^2\right)$, with the modified gauge couplings,
\begin{align}
    \overline{g}_s=g_s\left(1+C_{HG}v^2\right),~~~\overline{g}=g \left(1+C_{HW}v^2\right),~~~\overline{g}'=g' \left(1+C_{HB}v^2\right).
\end{align}
%
%In the SMEFT, these modified couplings are regarded as the gauge couplings in the broken phase.
%
In addition, upon removing kinetic mixing and diagonalizing the mass matrix, the photon is found to be massless, and the $W$ and $Z$ boson masses are given by,
\begin{align}
    M^2_W&=\frac{\overline{g}^2v^2}{4},\quad M^2_Z=\frac{v^2}{4}\left({\overline{g}'}^2+\overline{g}^2\right)+\frac{1}{8}v^4 C_{HD} \left({\overline{g}'}^2+\overline{g}^2\right)+\frac{1}{2}v^4 {\overline{g}'}\overline{g} C_{HWB}.\label{eq:MW_SMEFT}
\end{align}
Combining these expressions, the gauge couplings in the unbroken phase can be written in terms of physical observables and SMEFT Wilson coefficients at the electroweak symmetry breaking scale, which can be treated as input parameters.
In particular, we explicitly present the expression for $g$ at one loop level in subsection~\ref{sec:matching_loop}.
As will be explained, these physical parameters, including the gauge couplings and the SMEFT Wilson coefficients, are evaluated with RGE running in the unbroken phase. 
%So, for convenience, we listed the gauge couplings~\eqref{eq:gs}, \eqref{eq:g}, and \eqref{eq:g'}, defined in the unbroken phase.

\subsection{Matching at one-loop level}
\label{sec:matching_loop}
%
%We summarize the notation of the SMEFT (please also look at Appendix~\ref{sec:not}) to enhance our understanding of the forthcoming sections.
%

By improving the above considerations up to their one-loop corrections, additional SMEFT effects arise.
The particularly important $\overline{\rm MS}$ parameters of the Lagrangian are given by,
\begin{align}
    m_{\phi}^2 &=M^2_h +{\rm Re}\, \Pi_h(M^2_h),\label{eq:mphi}
    \\
    m^2_W &= M^2_W +{\rm Re}\, \Pi_W (M^2_W),\label{eq:mW}
    \\
    m^2_t &= M^2_t \left(1+2 {\rm Re}\, \Sigma_t(M^2_t)\right),\label{eq:mr}
    \\
    g^2 &=g_{0}^2 +\delta g^2,\label{eq:g2}
\end{align}
where $m_{\phi}, m_W$, and $m_t$ are the $\overline{{\rm MS}}$-masses of Higgs, $W$ boson, and top quark, respectively, $g$ is the $\overline{{\rm MS}}$-$SU(2)_L$ gauge coupling, $M_h, M_W$, and $M_t$ are pole masses of Higgs, $W$ boson, and top quark, respectively.
When the tree-level gauge coupling $g_{0}$ is determined from the measurement of $G_F$, $g_{0}$ is expressed as $g_{0}^2:= 4\sqrt{2}G_F M^2_W-8 M^2_W C_{HW0} +2M_W^2 \left[\left((C_{ll0})_{\mu ee\mu}+(C_{ll0})_{e\mu\mu e}\right)-2\left((C^{(3)}_{Hl0})_{ee}+(C^{(3)}_{Hl0})_{\mu\mu}\right)\right]$, where indices of 0 are used to explicitly indicate that the corresponding contributions arise at tree level.
Also, $\Pi_{h,W}$, and $\Sigma_t$ are one-loop self-energies, and $\delta g^2$ is one-loop correction to the $SU(2)_L$ gauge coupling.
%
%For brevity, we neglected one-loop corrections to the $U(1)_Y$ gauge coupling (also see Ref.~\cite{}).
%
The SMEFT operators listed in Table~\ref{tab:dim6} potentially contribute to the one-loop correction appearing on the right-hand side of eqs.~\eqref{eq:mphi}\text{--}\eqref{eq:g2} as follows:
\begin{align}
    \Pi_h(M^2_h)&=\Pi_{h,\,{\rm SM}}(M^2_h)+\Pi_{h,\,{\rm SMEFT}}(M^2_h),\label{eq:Pih}
    \\
    \Pi_W(M^2_W)&=\Pi_{W,\,{\rm SM}}(M^2_W)+\Pi_{W,\,{\rm SMEFT}}(M^2_W),
    \\
    \Sigma_t(M^2_t)&=\Sigma_{\rm SM}(M_t^2)+\Sigma_{\rm SMEFT}(M_t^2),
    \\
    \frac{\delta g^2}{g_{0}^2}&=\left(\frac{\delta g^2}{g_{0}^2}\right)_{\rm SM}+\left(\frac{\delta g^2}{g_{0}^2}\right)_{\rm SMEFT},\label{eq:delg2}
\end{align}
where the indices SM and SMEFT indicate the contributions from the SM and the SMEFT, respectively.
The SM contributions~\cite{Kajantie:1995dw} are listed in Appendix~\ref{sec:SM_self}.
From eqs.~\eqref{eq:mphi}\text{--}\eqref{eq:g2}, the $\overline{{\rm MS}}$-parameters, {\it e.g.,} eqs.~\eqref{eq:m2_tree} and \eqref{eq:lambda_tree}, are improved at the one-loop level and are expressed in terms of the physical parameters.
See Appendix~\ref{sec:MSbar_para} for the details.
At the $\overline{\rm MS}$ scale $\overline{\mu}=M_Z$, the $\overline{\rm MS}$ parameters are matched at the one-loop level using eqs~\eqref{eq:mu_loop}\text{--}\eqref{eq:CuHloop}. 
These parameters are evolved from $\overline{\mu} = M_Z$ to a typical scale of the SFO-EWPT using the RGEs, taking into account the effects of dimension-six SMEFT operators.
The consequences of these SMEFT effects on the SFO-EWPT will be discussed in Section~\ref{sec:GW_sec}.

Now, we provide expressions of the one-loop SMEFT effects on eqs.~\eqref{eq:Pih}\text{--}\eqref{eq:delg2}, and eqs.~\eqref{eq:CHloop}\text{--}\eqref{eq:CuHloop}.
%
%
%This work focuses on the leading-log terms arising from the SMEFT one-loop corrections. 
%
In what follows, we will examine three categories of one-loop effects within the SMEFT. 
These categories include corrections induced by Higgs self-couplings, associated with $m^2$ and $\lambda$; corrections arising from gauge couplings, including $g_s$, $g$, and $g'$; and corrections induced by the top Yukawa coupling $Y_t$.
Focusing on the logarithmic terms, we list the corresponding corrections in Tables~\ref{tab:Higgs_self}\text{--}\ref{tab:delCuH} for the three types of self-energies: $\Pi_h$, $\Pi_W$, and $\Sigma_t$, as well as for the $SU(2)_L$ gauge coupling $g$, and the four SMEFT Wilson coefficients $c_{H,{\rm kin}}$, $C_H$, $C_{HW}$, and $C_{uH}$.
While the logarithmic terms are listed in these tables for reference, the SMEFT RGE running is evaluated numerically in the analysis of the SFO-EWPT.
See subsection~\ref{sec:proc} for the details of the procedures to evaluate the SFO-EWPT.
In these Tables, the SMEFT operator one-loop correction normalized by $\frac{(4\pi)^2}{\ln \left(\overline{\mu}/M_W\right)}\times  C_{\rm SMEFT}^{-1}$ is shown for each SMEFT Wilson coefficient $C_{\rm SMEFT}$.
We showed these corrections by introducing parameters $h:=M_h/M_W, t:=M_t/M_W, z:=M_Z/M_W$, and $s:=g_s/\left(2M_W/v_0\right)$.
As studied in Ref.~\cite{Hashino:2022ghd}, although these one-loop corrections have a small effect on the Higgs potential, they can still be measurable in GW observations once the SFO-EWPT occurs, {\it e.g.}, via the SMEFT $(H^{\dagger}H)^3$ operator.
Prescriptions to measure these effects will be explained in Section~\ref{sec:GW_sec}.

    \begin{table}[htb]
  \begin{center}
    \begin{tabular}{|c|c|} \hline \hline
     Operator  & $\Pi_{h,{\rm SMEFT}}\left(M^2_h\right)\times \frac{(4\pi)^2}{\ln (\overline{\mu}/M_W)}\times \frac{1}{C_{\rm SMEFT}}$ 
     \\ \hline \hline
      $\mathcal{O}_H$ & $-36 M^2_W v_0^2 \frac{1}{h^2}\left(-2 +h^4 +4 t^4 -z^4\right)$ 
      \\ \hline                           
         $\mathcal{O}_{H\Box}$ & $4M_W^4 \left(14 h^4 + 6 \left(-2 +4 t^4 -z^4\right)+h^2 \left(6 t^2-3 (2+z^2)\right)\right)$ 
         \\ \hline                           
            $\mathcal{O}_{HD}$ & $-M^4_W \left(13 h^4 +3 h^2 \left(-2+2 t^2 +z^2\right)+12 \left(-1+2 t^4 -2 z^4\right)\right)$ 
            \\ \hline 
           $\mathcal{O}_{HWB}$ & $-24 M^4_W \sqrt{z^2-1} \left(h^2-3 z^2\right)$
           \\  \hline          
           $\mathcal{O}_{HB}$ & $-24 M^4_W \left(h^2-3 z^2\right)\left(z^2-1\right)$ 
           \\   \hline       
           $\mathcal{O}_{HW}$ & $72 M^4_W \left(2-h^2+z^2\right)$ 
           \\    \hline   
           $(\mathcal{O}_{uH})_{33}$ & $-12 \sqrt{2} M^3_W v_0 t\left(h^2-6 t^2 \right)$ 
           \\ \hline   
           $(\mathcal{O}_{ll})_{2112}$ & $3M_W^4 h^2 \left(-2+h^2+2 t^2-z^2\right)$ 
           \\ \hline   
           $(\mathcal{O}_{ll})_{1221}$ & $3M_W^4 h^2 \left(-2+h^2+2 t^2-z^2\right)$ 
           \\ \hline   
           $(\mathcal{O}_{Hl}^{(3)})_{11}$ & $-6 M^4_W h^2 \left(-2+h^2+2 t^2-z^2\right)$ 
           \\ \hline   
           $(\mathcal{O}_{Hl}^{(3)})_{22}$ & $-6 M^4_W h^2 \left(-2+h^2+2 t^2-z^2\right)$ 
           \\ 
   \hline  \hline  
    \end{tabular}
        \caption{The SMEFT one-loop corrections to the self energy of the Higgs field $\Pi_h\left(M_h^2\right)$.
        These corrections are shown by introducing parameters $h:=M_h/M_W, t:=M_t/M_W, z:=M_Z/M_W$, and $s:=g_s/\left(2M_W/v_0\right)$.
        The first column denotes the SMEFT dimension-six operator contributing to $\Pi_h\left(M_h^2\right)$. 
        The second column shows the correction to $\Pi_h\left(M_h^2\right)$ induced by the SMEFT Wilson coefficient $C_{\rm SMEFT}$, which corresponds to the operator listed in the first column.
        The correction is normalized by $\frac{(4\pi)^2}{\ln \left(\overline{\mu}/M_W\right)}\times  C_{\rm SMEFT}^{-1}$, where $\overline{\mu}$ denotes the $\overline{\rm MS}$ scale.
        Only the leading logarithmic terms are listed here for reference.
        }
 \label{tab:Higgs_self}
  \end{center}
\end{table}
    \begin{table}[htb]
  \begin{center}
    \begin{tabular}{|c|c|} \hline \hline
     Operator  & $\Pi_{W,{\rm SMEFT}}\left(M^2_W\right)\times \frac{(4\pi)^2}{\ln (\overline{\mu}/M_W)}\times \frac{1}{C_{\rm SMEFT}}$ 
     \\ \hline \hline
      $\mathcal{O}_H$ & $12 M^2_W v_0^2$ 
      \\ \hline                           
         $\mathcal{O}_{H\Box}$ & $-\frac{4}{3}M^4_W \frac{1}{h^2} \left(-10 h^2 + 15 h^4 +18 \left(2-4 t^4 +z^4\right)\right)$ 
         \\ \hline                           
            $\mathcal{O}_{HD}$ & $M^4_W \frac{1}{h^2}\left(12+5 h^4 -24 t^4 +3 h^2 \left(-4+z^2\right)\right)$ 
            \\ \hline 
           $\mathcal{O}_{HWB}$ & $8 M^4_W\frac{1}{h^2}\left(h^2-3 z^2\right)\sqrt{-1+z^2}$ 
           \\  \hline          
           $\mathcal{O}_{HB}$ & $-24 M^4_W \frac{z^2 }{h^2}\left(-1+z^2\right)$ 
           \\   \hline       
           $\mathcal{O}_{HW}$ & $-8 M^4_W \left(-2+\frac{12}{h^2}+h^2 -\frac{12t^4}{h^2}+\frac{3z^2}{h^2}+\frac{3z^4}{h^2}\right)$ 
           \\    \hline   
           $(\mathcal{O}_{uH})_{33}$ & $-24 \sqrt{2} M^3_W v_0 \frac{t^3}{h^2}$ 
           \\ \hline   
           $(\mathcal{O}_{Hl}^{(3)})_{11}$ & $\frac{2M_W^4}{3}\frac{1}{h^2}\left(9h^4 +h^2 \left(28 +18 t^2 -9 z^2\right)+18 \left(2-4t^4 +z^4\right)\right)$ 
           \\ \hline   
        $(\mathcal{O}_{Hl}^{(3)})_{22}$ & $\frac{2M_W^4}{3}\frac{1}{h^2}\left(9h^4 +h^2 \left(28 +18 t^2 -9 z^2\right)+18 \left(2-4t^4 +z^4\right)\right)$ 
           \\ \hline   
           $(\mathcal{O}_{Hl}^{(3)})_{33}$ & $\frac{16}{3}M^4_W$ 
           \\ \hline   
           $(\mathcal{O}_{Hq}^{(3)})_{11}$ & $16 M^4_W$ 
           \\ \hline   
           $(\mathcal{O}_{Hq}^{(3)})_{22}$ & $16 M^4_W$ 
           \\ \hline   
           $(\mathcal{O}_{Hq}^{(3)})_{33}$ & $16 M^4_W \left(1-\frac{3}{2}t^2\right)$ 
           \\ \hline   
           $\mathcal{O}_{W}$ & $-\frac{240 M^5_W}{v_0}$ 
           \\ \hline   
           $(\mathcal{O}_{uW})_{33}$ & $-24 \sqrt{2} M^4_W t$ 
           \\ \hline   
           $(\mathcal{O}_{ll})_{2112}$ & $-\frac{M_W^4}{3}\frac{1}{h^2}\left(9 h^4 + h^2 (20 + 18 t^2 - 9 z^2) + 18 (2 - 4 t^4 + z^4)\right)$ 
           \\ \hline   
           $(\mathcal{O}_{ll})_{1221}$ & $-\frac{M^4_W}{3}\frac{1}{h^2}\left(9 h^4 + h^2 (20 + 18 t^2 - 9 z^2) + 18 (2 - 4 t^4 + z^4)\right)$
   \\ \hline  
    \end{tabular}
        \caption{The same table as Table~\ref{tab:Higgs_self}, but for the $W$ boson self energy $\Pi_W \left(M^2_W\right)$.
        }
 \label{tab:W_self}
  \end{center}
\end{table}
    \begin{table}[htb]
  \begin{center}
    \begin{tabular}{|c|c|} \hline \hline
     Operator  & $\Sigma_{{\rm SMEFT}}\left(M^2_t\right)\times \frac{(4\pi)^2}{\ln (\overline{\mu}/M_W)}\times \frac{1}{C_{\rm SMEFT}}$ 
     \\ \hline \hline
      $\mathcal{O}_H$ & $6 v_0^2$ 
      \\ \hline                           
         $\mathcal{O}_{H\Box}$ & $-2M^2_W \frac{1}{h^2}\left(5h^4 - 3 h^2 t^2 +6 \left(2-4 t^4 +z^4\right)\right)$ 
         \\ \hline                           
            $\mathcal{O}_{HD}$ & $\frac{M^2_W}{6}\frac{1}{h^2}\left(36 + 15 h^4 -72 t^4 + h^2 \left(-6 t^2+8 z^2\right)\right)$ 
            \\ \hline 
           $\mathcal{O}_{HWB}$ & $\frac{4}{3}M^2_W \frac{1}{h^2}\left(10 h^2-9 z^2\right)\sqrt{-1+z^2}$ 
           \\  \hline          
           $\mathcal{O}_{HB}$ & $-12 M^2_W \frac{1}{h^2}z^2 \left(-1+z^2\right)$ 
           \\   \hline       
           $\mathcal{O}_{HW}$ & $-12 M^2_W \frac{1}{h^2}\left(2+z^2\right)$
           \\    \hline   
           $(\mathcal{O}_{uH})_{33}$ & $3\sqrt{2} M_W v_0 \frac{1}{h^2 t}\left(2-h^2 t^2 -8 t^4 +z^4\right)$ 
           \\ \hline   
           $(\mathcal{O}_{Hu})_{33}$ & $-2 M^2_W \left(-4+t^2+z^2\right)$ 
           \\ \hline   
            $(\mathcal{O}_{Hq}^{(1)})_{33}$ & $2M^2_W \left(4+t^2-4z^2\right)$ 
           \\ \hline  $(\mathcal{O}_{Hq}^{(3)})_{33}$ & $8 M^2_W \left(-1+z^2\right)$ 
           \\ \hline   
           $(\mathcal{O}_{uW})_{33}$ & $2\sqrt{2} M^2_W \frac{1}{t}\left(14-5 z^2\right)$ 
            \\ \hline   
           $(\mathcal{O}_{qu}^{(1)})_{3333}$ & $-8 M^2_W t^2$ 
           \\ \hline   
           $(\mathcal{O}_{qu}^{(8)})_{3333}$ & $-\frac{32}{3} M^2_W t^2$ 
           \\ \hline   
           $\mathcal{O}_{HG}$ & $-64 M_W^2 s^2$ 
           \\ \hline   
           $\mathcal{O}_{H\widetilde{G}}$ & $-64 i M_W^2 s^2$ 
           \\ \hline   
           $\mathcal{O}_{H\widetilde{W}}$ & $-18 i M^2_W$ 
           \\ \hline   
           $\mathcal{O}_{H\widetilde{B}}$ & $-\frac{34}{3}i M^2_W \left(-1+z^2\right)$ 
           \\ \hline   
           $\mathcal{O}_{H\widetilde{W}B}$ & $2i M^2_W \sqrt{-1+z^2}$ 
           \\ \hline   
           $\left(\mathcal{O}_{uB}\right)_{33}$ & $2\sqrt{2} M^2_W \frac{1}{t}\sqrt{-1+z^2} \left(-8+5 t^2 + 5z^2\right)$ 
           \\ \hline   
           $\left(\mathcal{O}_{uG}\right)_{33}$ & $32 \sqrt{2} M_W^2 s t$ 
           \\ \hline   
           $\left(\mathcal{O}_{ll}\right)_{1221}$ & $-\frac{M_W^2}{6}\frac{1}{h^2}\left(9 h^4 + h^2 (-8  - 9 t^2 + 8 z^2) + 18 \left(2 - 4 t^4 + z^4\right)\right)$ 
           \\ \hline   
           $\left(\mathcal{O}_{ll}\right)_{2112}$ & $-\frac{M_W^2}{6}\frac{1}{h^2}\left(9 h^4 + h^2 (-8  - 9 t^2 + 8 z^2) + 18 \left(2 - 4 t^4 + z^4\right)\right)$ 
           \\ \hline   
           $\left(\mathcal{O}_{Hl}^{(3)}\right)_{11}$ & $\frac{M_W^2}{3}\frac{1}{h^2}\left(9 h^4 + h^2 \left(-8  - 9 t^2 + 8 z^2\right) + 18 \left(2 - 4 t^4 + z^4\right)\right)$ 
           \\ \hline   
           $\left(\mathcal{O}_{Hl}^{(3)}\right)_{22}$ & $\frac{M_W^2}{3}\frac{1}{h^2}\left(9 h^4 + h^2 \left(-8  - 9 t^2 + 8 z^2\right) + 18 \left(2 - 4 t^4 + z^4\right)\right)$
           \\ 
   \hline  \hline  
    \end{tabular}
        \caption{The same table as Table~\ref{tab:Higgs_self}, but for the top quark self energy $\Sigma_t \left(M^2_t\right)$.
        }
 \label{tab:top_self}
  \end{center}
\end{table}
  \begin{table}[htb]
  \begin{center}
    \begin{tabular}{|c|c|} \hline \hline
     Operator  & $\left(\delta g^2/g_0^2\right)_{\rm SMEFT} \times \frac{(4\pi)^2}{\ln (\overline{\mu}/M_W)}\times \frac{1}{C_{\rm SMEFT}}$ 
     \\ \hline \hline                                $\mathcal{O}_{HW}$ & $\frac{8}{3}M_W^2\left(19-3 h^2\right)$ 
           \\ \hline   
         $\left(\mathcal{O}_{ll}\right)_{1221}$ & $-\frac{38}{3}M_W^2$ 
           \\ \hline   
           $\left(\mathcal{O}_{ll}\right)_{2112}$ &  $-\frac{38}{3}M_W^2$  
           \\ \hline   
           $\left(\mathcal{O}_{Hl}^{(3)}\right)_{11}$ & $\frac{76}{3}M_W^2$ 
           \\ \hline   
           $\left(\mathcal{O}_{Hl}^{(3)}\right)_{22}$ & $\frac{76}{3}M_W^2$
           \\ 
   \hline  \hline  
    \end{tabular}
        \caption{The same table as Table~\ref{tab:Higgs_self}, but for $\delta g^2/g_0^2$.
        }
 \label{tab:delg2}
  \end{center}
\end{table}
  \begin{table}[htb]
  \begin{center}
    \begin{tabular}{|c|c|} \hline \hline
     Operator  & $\delta c_{H,{\rm kin}}  \times \frac{(4\pi)^2}{\ln (\overline{\mu}/M_W)}\times \frac{1}{C_{\rm SMEFT}}$ 
     \\ \hline \hline   
          $\mathcal{O}_{H\Box}$ & $2M^2_W \frac{1}{h^2} \left(-12 + 3 h^4 + 24 t^4 - 6 z^4 + h^2 (4 + 6 t^2 - 3 z^2)\right)$ 
           \\ \hline  
           $\mathcal{O}_{HD}$ & $3 M^2_W \frac{1}{h^2} \left(2 - 4 t^4 + z^4 - h^2 \left(5 + t^2 - 2 z^2\right)\right)$ 
           \\ \hline   $\left(\mathcal{O}_{Hl}^{(3)}\right)_{11}$ & $8 M_W^2$ 
           \\ \hline   
           $\left(\mathcal{O}_{Hl}^{(3)}\right)_{22}$ & $8 M_W^2$ 
           \\ \hline  
           $\left(\mathcal{O}_{Hl}^{(3)}\right)_{33}$ & $8 M_W^2$ 
           \\ \hline  
           $\left(\mathcal{O}_{Hq}^{(3)}\right)_{11}$ & $24 M_W^2$ 
           \\ \hline  
           $\left(\mathcal{O}_{Hq}^{(3)}\right)_{22}$ & $24 M_W^2$ 
           \\ \hline  
           $\left(\mathcal{O}_{Hq}^{(3)}\right)_{33}$ & $24 M_W^2 \left(1-\frac{3}{2}t^2\right)$ 
           \\ 
   \hline  \hline  
    \end{tabular}
        \caption{The same table as Table~\ref{tab:Higgs_self}, but for $c_{H,{\rm kin}}$.
        }
 \label{tab:delCHkin}
  \end{center}
\end{table}
  \begin{table}[htb]
  \begin{center}
    \begin{tabular}{|c|c|} \hline \hline
     Operator  & $\delta C_{H}  \times \frac{(4\pi)^2}{\ln (\overline{\mu}/M_W)}\times \frac{1}{C_{\rm SMEFT}}$ 
     \\ \hline \hline 
          $\mathcal{O}_{H}$ & $\frac{18M^2_W}{v_0^2} \left(-2 + 3 h^2 + 2 t^2 - z^2\right)$ 
           \\ \hline  
          $\mathcal{O}_{H\Box}$ & $\frac{40 M_W^4}{3v_0^4}h^2 \left(2-3 h^2\right)$
           \\ \hline  
           $\mathcal{O}_{HD}$ & $\frac{12 M_W^4}{v_0^4}\left(h^4 - z^4 + h^2 (-2 + z^2)\right)$ 
           \\ \hline  
           $\mathcal{O}_{HWB}$ & $\frac{24 M_W^4}{v_0^4}\left(h^2-2 z^2\right)\sqrt{-1+z^2}$ 
           \\ \hline  
                      $\mathcal{O}_{HB}$ & $\frac{24 M^4_W}{v_0^4}(h^2 - 2 z^2) (-1 + z^2)$ 
           \\ \hline   
            $\mathcal{O}_{HW}$ & $\frac{24 M^4_W}{v_0^4}(3 h^2 - 2 (2 + z^2))$ 
           \\ \hline   
           $\left(\mathcal{O}_{uH}\right)_{33}$ & $\frac{12\sqrt{2} M^3_W}{v_0^3}t(h^2-4 t^2)$ 
           \\ \hline  $\left(\mathcal{O}_{Hl}^{(3)}\right)_{11}$ & $\frac{32}{3}\frac{M^4_W}{v_0^4}h^2$ 
           \\ \hline   
           $\left(\mathcal{O}_{Hl}^{(3)}\right)_{22}$ & $\frac{32}{3}\frac{M^4_W}{v_0^4}h^2$ 
           \\ \hline  
           $\left(\mathcal{O}_{Hl}^{(3)}\right)_{33}$ & $\frac{32}{3}\frac{M^4_W}{v_0^4}h^2$ 
           \\ \hline  
           $\left(\mathcal{O}_{Hq}^{(3)}\right)_{11}$ & $\frac{32 M_W^4}{v_0^4}h^2$ 
           \\ \hline  
           $\left(\mathcal{O}_{Hq}^{(3)}\right)_{22}$ & $\frac{32 M_W^4}{v_0^4}h^2$
           \\ \hline  
           $\left(\mathcal{O}_{Hq}^{(3)}\right)_{33}$ & $\frac{32 M_W^4}{v_0^4}h^2\left(1-\frac{3}{2}t^2\right)$ 
           \\ 
   \hline  \hline  
    \end{tabular}
        \caption{The same table as Table~\ref{tab:Higgs_self}, but for $C_H$.
        }
 \label{tab:delCH}
  \end{center}
\end{table}
  \begin{table}[htb]
  \begin{center}
    \begin{tabular}{|c|c|} \hline \hline
     Operator  & $\delta C_{HW}  \times \frac{(4\pi)^2}{\ln (\overline{\mu}/M_W)}\times \frac{1}{C_{\rm SMEFT}}$ 
     \\ \hline \hline 
           $\mathcal{O}_{HWB}$ & $\frac{4 M_W^2}{v_0^2} \sqrt{-1+z^2}$ 
           \\ \hline  
            $\mathcal{O}_{HW}$ & $\frac{2M_W^2}{3 v_0^2}\left(-44 + 9 h^2 +18 t^2 -9 z^2\right)$ 
            \\ \hline  
            $\mathcal{O}_{W}$ & $-\frac{120 M_W^3}{v_0^3}$ 
            \\ \hline  
            $\left(\mathcal{O}_{uW}\right)_{33}$ & $-\frac{12\sqrt{2}M_W^2}{v_0^2}t$ 
           \\ 
   \hline  \hline  
    \end{tabular}
        \caption{The same table as Table~\ref{tab:Higgs_self}, but for $C_{HW}$.
        }
 \label{tab:delCHW}
  \end{center}
\end{table}
  \begin{table}[htb]
  \begin{center}
    \begin{tabular}{|c|c|} \hline \hline
     Operator  & $\delta C_{uH}  \times \frac{(4\pi)^2}{\ln (\overline{\mu}/M_W)}\times \frac{1}{C_{\rm SMEFT}}$ 
     \\ \hline \hline 
          $\mathcal{O}_{H\Box}$ & $-\frac{2\sqrt{2}M_W^3}{3 v^3_0} t\left(-20 + 3h^2+18 t^2\right)$
           \\ \hline  
           $\mathcal{O}_{HD}$ & $\frac{\sqrt{2}M_W^3}{v^3_0}t \left(h^2+2 \left(-6+t^2+3z^2\right)\right)$
           \\ \hline  
           $\mathcal{O}_{HWB}$ & $-\frac{4\sqrt{2}M_W^3}{v^3_0} t\sqrt{-1+z^2}$ 
           \\ \hline  
                      $\mathcal{O}_{HB}$ & $\frac{68 \sqrt{2}M_W^3}{3v^3_0} t\left(-1+z^2\right)$
           \\ \hline   
            $\mathcal{O}_{HW}$ & $\frac{36\sqrt{2}M_W^3}{v^3_0}t$
           \\ \hline   
           $\left(\mathcal{O}_{uG}\right)_{33}$ & $-\frac{128 M_W^3}{v^3_0}s t^2$ 
           \\ \hline 
           $\left(\mathcal{O}_{uB}\right)_{33}$ & $-\frac{8 M_W^3}{v^3_0}\sqrt{-1+z^2} \left(-8 + 5t^2+5 z^2\right)$ 
           \\ \hline   
           $\left(\mathcal{O}_{uW}\right)_{33}$ & $\frac{8 M_W^3}{v^3_0} \left(-14 +5 z^2\right)$ 
           \\  \hline  
           $\left(\mathcal{O}_{uH}\right)_{33}$ & $\frac{M_W^2}{3v^2_0}\left(-46 + 36 h^2 - 96 s^2 + 153 t^2 - 35 z^2\right)$ 
           \\ \hline  
            $\mathcal{O}_{HG}$ & $\frac{128 \sqrt{2}M^3_W}{v^3_0}s^2 t$
           \\ \hline  
           $\mathcal{O}_{H\widetilde{G}}$ & $i\frac{128\sqrt{2}M_W^3}{v^3_0}s^2 t$
           \\ \hline  
            $\mathcal{O}_{H\widetilde{B}}$ & $i\frac{68\sqrt{2}M_W^3}{3v^3_0} t\left(-1+z^2\right)$
           \\ \hline  
           $\mathcal{O}_{H\widetilde{W}}$ & $i\frac{36\sqrt{2}M_W^3}{v^3_0}t$
           \\ \hline 
           $\mathcal{O}_{H\widetilde{W}B}$ & $-i \frac{4\sqrt{2}M_W^3}{v^3_0}t\sqrt{-1+z^2}$
           \\ \hline  $\left(\mathcal{O}_{Hl}^{(3)}\right)_{11}$ & $\frac{16\sqrt{2}M^3_W}{3v^3_0}t$
           \\ \hline   
           $\left(\mathcal{O}_{Hl}^{(3)}\right)_{22}$ & $\frac{16\sqrt{2}M^3_W}{3v^3_0}t$ 
           \\ \hline  
           $\left(\mathcal{O}_{Hl}^{(3)}\right)_{33}$ & $\frac{16\sqrt{2}M^3_W}{3v^3_0}t$
           \\ \hline  
           $\left(\mathcal{O}_{Hq}^{(1)}\right)_{33}$ & $-\frac{2\sqrt{2}M_W^3}{v^3_0}t \left(8+h^2 +2 t^2-8 z^2\right)$
           \\ \hline  $\left(\mathcal{O}_{Hq}^{(3)}\right)_{11}$ & $\frac{16\sqrt{2}M^3_W}{v^3_0}t$
           \\ \hline  
           $\left(\mathcal{O}_{Hq}^{(3)}\right)_{22}$ & $\frac{16\sqrt{2}M^3_W}{v^3_0}t$
           \\ \hline  
           $\left(\mathcal{O}_{Hq}^{(3)}\right)_{33}$ & $\frac{2\sqrt{2}M_W^3}{v^3_0}t \left(16 + 3 h^2 - 12 t^2 - 8 z^2\right)$
           \\ \hline  
           $\left(\mathcal{O}_{Hu}\right)_{33}$ & $\frac{2\sqrt{2}M_W^3}{v^3_0}t \left(h^2 +2 \left(-4+t^2+z^2\right)\right)$ 
           \\ \hline  
           $\left(\mathcal{O}_{qu}^{(1)}\right)_{3333}$ & $\frac{4\sqrt{2}M_W^3}{v^3_0}t \left(-h^2 +4 t^2\right)$
           \\ \hline  
           $\left(\mathcal{O}_{qu}^{(8)}\right)_{3333}$ & $\frac{16\sqrt{2}M_W^3}{3v^3_0}t \left(-h^2+4t^2\right)$
           \\ 
   \hline  \hline  
    \end{tabular}
        \caption{The same table as Table~\ref{tab:Higgs_self}, but for $C_{uH}$.
        }
 \label{tab:delCuH}
  \end{center}
\end{table}

\subsection{One-loop CW potential}
\label{sec:CW}
%
%As explained in subsection~\ref{sec:matching_loop}, physical parameters in the Higgs potential are modified by loop corrections involving the SMEFT operators.
%
At the one-loop level, the Coleman-Weinberg (CW) potential~\cite{Coleman:1973jx} also contributes to the effective Higgs potential.
The SMEFT operators can induce modifications to the CW potential, potentially impacting GW observations.
The CW contributions are given by:
\begin{align}
    V_{\rm CW}=\sum_{i}\frac{n_i}{64\pi^2}M_i^4\left(\phi\right) \left(
    \ln \left(\frac{M^2_i \left(\phi\right)}{\overline{\mu}^2}\right)-c_i
    \right),\label{eq:CW}
\end{align}
with the $\overline{\rm MS}$ scale $\overline{\mu}$.
As will be detailed in the subsection~\ref{sec:proc}, the right-hand side is evaluated at a typical scale of the SFO-EWPT.
The summation is conducted over the index $i$, which represents the Higgs boson $h$, the Nambu-Goldstone (NG) boson modes, as well as the $W$ and $Z$ bosons, and the top quark $t$.
The prefactor $n_i$ denotes the degrees of freedom of the modes ($n_h=1$, $n_{1+}=n_{1-}=1$, $n_{2+}=n_{2-}=2$, $n_W=6$, $n_Z=3$, and $n_t=-12$), and $c_i=5/6$ ($3/2$) for the gauge boson (other modes).
%
\begin{comment}
\begin{align}
    V_{\rm CW}=&-12 \frac{M_t^4(\phi)}{64 \pi^2}\left(
    \ln \left(\frac{M_t^2(\phi)}{\overline{\mu}^2}\right)-\frac{3}{2}
    \right)\notag
    \\
    &+6 \frac{M_W^4(\phi)}{64 \pi^2}\left(
    \ln \left(\frac{M_W^2(\phi)}{\overline{\mu}^2}\right)-\frac{5}{6}
    \right)+3 \frac{M_Z^4(\phi)}{64 \pi^2}\left(
    \ln \left(\frac{M_Z^2(\phi)}{\overline{\mu}^2}\right)-\frac{5}{6}
    \right)\notag 
    \\
    &+ \frac{M_{\phi}^4(\phi)}{64\pi^2} \left(
    \ln \left(\frac{M_{\phi}^2(\phi)}{\overline{\mu}^2}\right)-\frac{3}{2}
    \right)+ 2 \sum_{i=\pm}\frac{(M_{2}^i)^{4}(\phi)}{64\pi^2} \left(
    \ln \left(\frac{(M_{2}^i)^2(\phi)}{\overline{\mu}^2}\right)-\frac{3}{2}
    \right)\notag
    \\
    &+\sum_{i=\pm}\frac{(M_{1}^i)^{4}(\phi)}{64\pi^2} \left(
    \ln \left(\frac{(M_{1}^i)^2(\phi)}{\overline{\mu}^2}\right)-\frac{3}{2}
    \right),\label{eq:CW}
\end{align}
\end{comment}
%
Incorporating the contributions from the SMEFT dimension-six operators, we have calculated the field-dependent masses in eq.~\eqref{eq:CW}. 
The detailed calculations and results are summarized in Appendix~\ref{sec:FDM}.
%
%
%In section~\ref{sec:proc}, we will study the EWPT based on the above building blocks.

\begin{figure}[t]
\begin{center}
\includegraphics[width=15.cm]{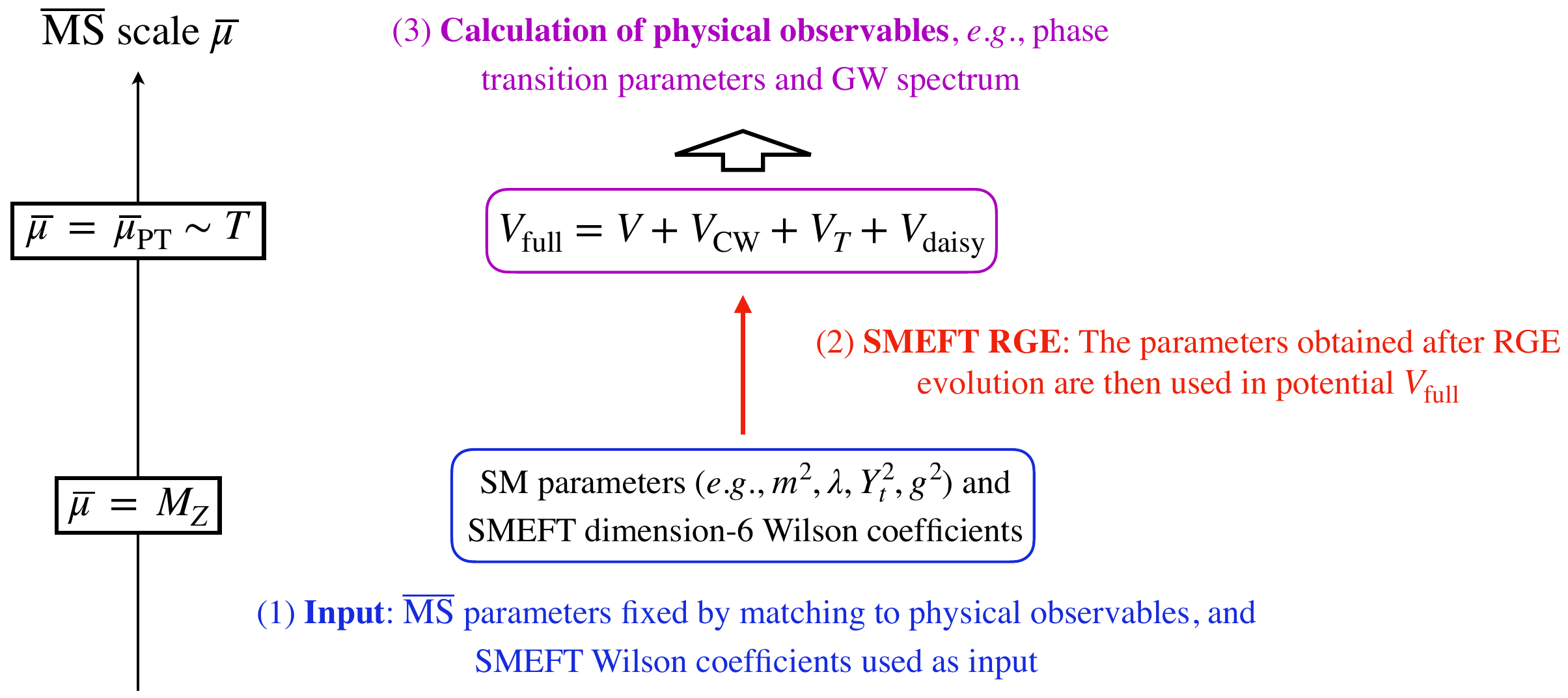}
\end{center}
\vspace{-0.4cm}
\caption{A schematic illustration of the procedures for analyzing the SFO-EWPT under the SMEFT RGE effects.
The procedures are mainly classified into three categories: {\bf (1)} At the $\overline{\rm MS}$ scale $\overline{\mu} = M_Z$, the one-loop improved $\overline{\rm MS}$ parameters are determined by physical observables and the input values of the SMEFT Wilson coefficients, {\bf (2)} the $\overline{\rm MS}$ parameters are evaluated at typical scales of the SFO-EWPT, $\overline{\mu} = \overline{\mu}_{\rm PT} \sim T$, using the one-loop SMEFT RGE running, and then substituted into the effective potential $V_{\rm full}$ in eq.~\eqref{eq:Vfull}, and {\bf (3)} physical observables, such as phase transition parameters and the GW spectrum, are calculated from $V_{\rm full}$.
As explained in subsection~\ref{sec:GW}, we choose the nucleation temperature $T_n$ as the typical scale of the SFO-EWPT and evaluate the renormalization scale uncertainties by considering two scales: $\overline{\mu}_{\rm PT} = 2\pi T_n$ and $T_n/2$.
}
\label{fig:sch}
\end{figure}

\subsection{One-loop thermal effective potential}
\label{sec:thermal}
Next, we provide the finite temperature effects on the Higgs potential. 
The thermal loop effects  on the Higgs effective potential are expressed as follows:
\begin{align}
    V_T=\frac{T^4}{2\pi^2}
    \left\{
    \sum_i n_i I_{\rm B} \left(\left(M_i(\phi)/T\right)^2 \right)
    + n_t I_{\rm F} \left(\left(M_t(\phi)/T\right)^2 \right)
    \right\},\label{eq:VT}
\end{align}
where the degrees of freedom associated with the modes $n_i$ and the field-dependent mass $ M^2_i(\phi)$ are consistent with the expression provided in eq.~\eqref{eq:CW}.
The thermal functions~\cite{Dolan:1973qd} are given as
\begin{align}
    I_{\rm B}(a_i^2)&=\int_0^{\infty} dx x^2 \ln \left[1-{\rm exp}\left(-\sqrt{x^2 +a_i^2}\right)\right],
    \\
    I_{\rm F}(a_i^2)&=\int_0^{\infty}dx x^2 \ln \left[
    1+{\rm exp}\left(-\sqrt{x^2 +a_i^2}\right)
    \right],
\end{align}
with $a_i:=M_i(\phi)/T$.
In this paper, we adopt the conventional daisy-resummation method to examine the impact of renormalization scale uncertainties~\cite{Croon:2020cgk} in the Fisher matrix analysis.
To this end, and to avoid IR divergences at high temperatures, we include the following daisy diagrams in the effective potential~\cite{Parwani:1991gq,Arnold:1992rz}:
\begin{align}
    V_{\rm daisy}=\frac{T}{12\pi}\sum_{i} n'_i
    \left(
    \left[M^2_{i,{\rm res.}}\left(\phi,0\right)\right]^{3/2}-
    \left[M^2_{i,{\rm res.}}\left(\phi,T\right)\right]^{3/2}
    \right),\label{eq:daisy}
\end{align}
where $n'_i$ denotes the degrees of freedom of the modes ($n'_h=1$, $n'_{1+}=n'_{1-}=1$, $n'_{2+}=n'_{2-}=2$, $n'_W=2$, $n'_Z=1$, and $n'_{A'}=1$).
In Appendix~\ref{sec:resum}, we provide the thermally resummed masses with the SMEFT dimension-six operator effects.
See also Ref.~\cite{Croon:2020cgk} for further details.

\subsection{Procedures to evaluate the SFO-EWPT}
\label{sec:proc}
So far, we have presented the key components of the effective potential.
This subsection outlines our evaluation procedures for the SFO-EWPT to facilitate understanding of the following sections.
Further details can also be found in Ref.~\cite{Croon:2020cgk}.
As explained, the SFO-EWPT is described by the effective potential, which consists of the following four terms:
\begin{align}
    V_{\rm full}(\phi,T)= V +V_{\rm CW} + V_T +V_{\rm daisy},\label{eq:Vfull}
\end{align}
where the first term, denoted as $V$, represents the tree-level Higgs potential as outlined in eq.~\eqref{eq:Vtree}, the second term, $V_{\rm CW}$, corresponds to the one-loop CW potential~\eqref{eq:CW}, the third term, $V_T$, captures the thermal one-loop potential~\eqref{eq:VT}, while the final term, $V_{\rm daisy}$, accounts for the contributions from daisy diagrams~\eqref{eq:daisy}.
The effective potential~\eqref{eq:Vfull} is evaluated at the typical energy scale of the SFO-EWPT $\overline{\mu}=\overline{\mu}_{\rm PT}$.
Following Ref.~\cite{Croon:2020cgk}, we examine two different scales, $\overline{\mu}_{\rm PT} = T/2$ and $2\pi T$, where $T$ is chosen as the nucleation temperature $T_n$, as explained in subsection~\ref{sec:GW}.
Since $\overline{\mu}_{\rm PT}$ is not fixed between $T/2$ and $2\pi T$, this choice of the renormalization scale introduces a potential theoretical uncertainty.
In our analysis, the numerical values of the SMEFT Wilson coefficients, as presented in Table~\ref{tab:dim6}, are input at the scale $\overline{\mu} = M_Z$ and then evaluated at the scale $\overline{\mu} =\overline{\mu}_{\rm PT}$ through one-loop RGE running.

For clarity, Fig.~\ref{fig:sch} schematically illustrates the procedure for evaluating the SFO-EWPT.
The procedure can be divided into three main steps: {\bf (1)} The $\overline{\rm MS}$ parameters in eq.~\eqref{eq:Vfull} at the $\overline{\rm MS}$ scale $\overline{\mu}=M_Z$ are determined from physical observables and the input values of the SMEFT Wilson coefficients. 
{\bf (2)} These parameters are then evaluated at the scale $\overline{\mu}=\overline{\mu}_{\rm PT}$ via one-loop SMEFT RGE running and substituted into eq.~\eqref{eq:Vfull}. 
In this work, we evaluate eq.~\eqref{eq:Vfull} at two different scales, $\overline{\mu}_{\rm PT}=T/2$ and $2\pi T$, to examine the theoretical uncertainty arising from renormalization scale dependencies. 
{\bf (3)} As explained in subsection~\ref{sec:GW}, physical observables—such as phase transition parameters and the GW spectrum—are calculated using eq.~\eqref{eq:Vfull}, evaluated at the scale $\overline{\mu} = \overline{\mu}_{\rm PT}$.

We now list how the effects of the SMEFT dimension-six operators are incorporated into the four terms in eq.~\eqref{eq:Vfull}, one by one, as follows:

\begin{itemize}
    \item {\bf The effects of the SMEFT dimension-six operators on $V$} arise from the tree-level corrections shown in eqs.~\eqref{eq:Vtree}, \eqref{eq:mu_loop}, and \eqref{eq:lambda_loop}, as well as from the one-loop SMEFT RGE running of the parameters $m^2$, $\lambda$, and $C_H$ in eq.~\eqref{eq:Vtree}, from the scale $\overline{\mu} = M_Z$ to $\overline{\mu} = \overline{\mu}_{\rm PT}$, where $\overline{\mu}_{\rm PT}$ denotes the typical energy scale of the SFO-EWPT.
Using eqs.~\eqref{eq:mu_loop}, \eqref{eq:lambda_loop}, and \eqref{eq:CHloop}, the SMEFT operator effects are incorporated into the parameters in $V$ at $\overline{\mu} = M_Z$, which are then evolved to $\overline{\mu} = \overline{\mu}_{\rm PT}$ via the one-loop SMEFT RGE.
In our analysis, we mainly focus on the theoretical uncertainty in the GW spectrum arising from the choice of the renormalization scale $\overline{\mu} = \overline{\mu}_{\rm PT}$.

    To enhance clarity, we list the SMEFT operators relevant to the potential $V$.
The tree-level corrections arise from the following operators:
    \begin{align}
    \mathcal{O}_H&= \left(H^{\dagger}H\right)^3,
    \\
    \mathcal{O}_{H\Box}&=\left(H^{\dagger}H\right)\Box \left(H^{\dagger}H\right),
    \\
    \mathcal{O}_{HD}&=\left(H^{\dagger}D_{\mu}H\right)^{\ast} \left(H^{\dagger}D_{\mu}H\right),
    \\
    (\mathcal{O}_{ll})_{\mu ee\mu}&=\left(\bar{l}_2 \gamma_{\mu}l_1\right)\left(\bar{l}_1 \gamma^{\mu}l_2\right),
        \\
    (C_{Hl}^{(3)})_{11}&=\left(H^{\dagger}i\overline{D}^I_{\mu}H
    \right)\left(\bar{l}_1 \tau^I \gamma^{\mu}l_1\right),
    \\
    (C_{Hl}^{(3)})_{22}&=\left(H^{\dagger}i\overline{D}^I_{\mu}H
    \right)\left(\bar{l}_2 \tau^I \gamma^{\mu}l_2\right).    
    \end{align}
    The operator $C_H$ appears explicitly in eq.~\eqref{eq:Vtree}, while the other operators listed above contribute to the potential $V$ through eqs.~\eqref{eq:mu_loop} and \eqref{eq:lambda_loop}.

    On the other hand, one-loop level corrections to $V$ arise from the RGE effects on $m^2$, $\lambda$, and $C_H$. 
    In addition, they include contributions from the self-energies $\Pi_h(M_h^2)$ and $\Pi_W(M_W^2)$, corrections to the gauge coupling $\delta g^2 / g_0^2$, and SMEFT Wilson coefficients $\delta C_H / C_{H0}$, $\delta c_{H,{\rm kin}} / c_{H,{\rm kin}0}$, and $\delta C_{HW} / C_{HW0}$.
    For details, see Appendix~\ref{sec:MSbar_para}. % 
    These one-loop SMEFT corrections are summarized in Tables~\ref{tab:Higgs_self}–\ref{tab:delCHW} up to the leading logarithmic terms.

    \item {\bf The effects of dimension-six SMEFT operators on $V_{\rm CW}$} at one-loop level are encoded through the field-dependent masses given in eqs.~\eqref{eq:MW_phi}–\eqref{eq:M2_phi} and \eqref{eq:Mt_phi}. 
    The SMEFT operators contributing to these field-dependent masses are listed below:
    \begin{align}
    \mathcal{O}_H&=\left(H^{\dagger}H\right)^3,
    \\
    \mathcal{O}_{H\Box}&=\left(H^{\dagger}H\right)\Box\left(H^{\dagger}H\right),
    \\
    \mathcal{O}_{HD}&=\left(H^{\dagger}D_{\mu}H\right)^{\ast}\left(H^{\dagger}D_{\mu}H\right),
    \\
    \mathcal{O}_{HB}&=H^{\dagger}H B_{\mu\nu}B^{\mu\nu},
    \\
    \mathcal{O}_{HW}&=H^{\dagger}H W^I_{\mu\nu}W^{I\mu\nu},
    \\
    \mathcal{O}_{HWB}&=H^{\dagger}\tau^I H W^I_{\mu\nu} B^{\mu\nu},
    \\
    (\mathcal{O}_{uH})_{33}&=\left(H^{\dagger}H\right)\left(\overline{q}_3 u_3 \widetilde{H}\right).
    \end{align}
    In addition to the above effects, the CW potential also receives higher-loop corrections through the RGE running of the parameters appearing in the field-dependent masses given in eqs.~\eqref{eq:MW_phi}–\eqref{eq:M2_phi} and \eqref{eq:Mt_phi}.

    \item {\bf The effects of the SMEFT dimension-six operators on the finite-temperature contributions, $V_T$ and $V_{\rm daisy}$}, arise primarily through the field-dependent masses given in eqs.~\eqref{eq:MW_phi}–\eqref{eq:M2_phi} and \eqref{eq:Mt_phi}, which enter the thermally resummed masses.

\end{itemize}

\begin{figure}[t]
\begin{center}
\includegraphics[width=8.cm]{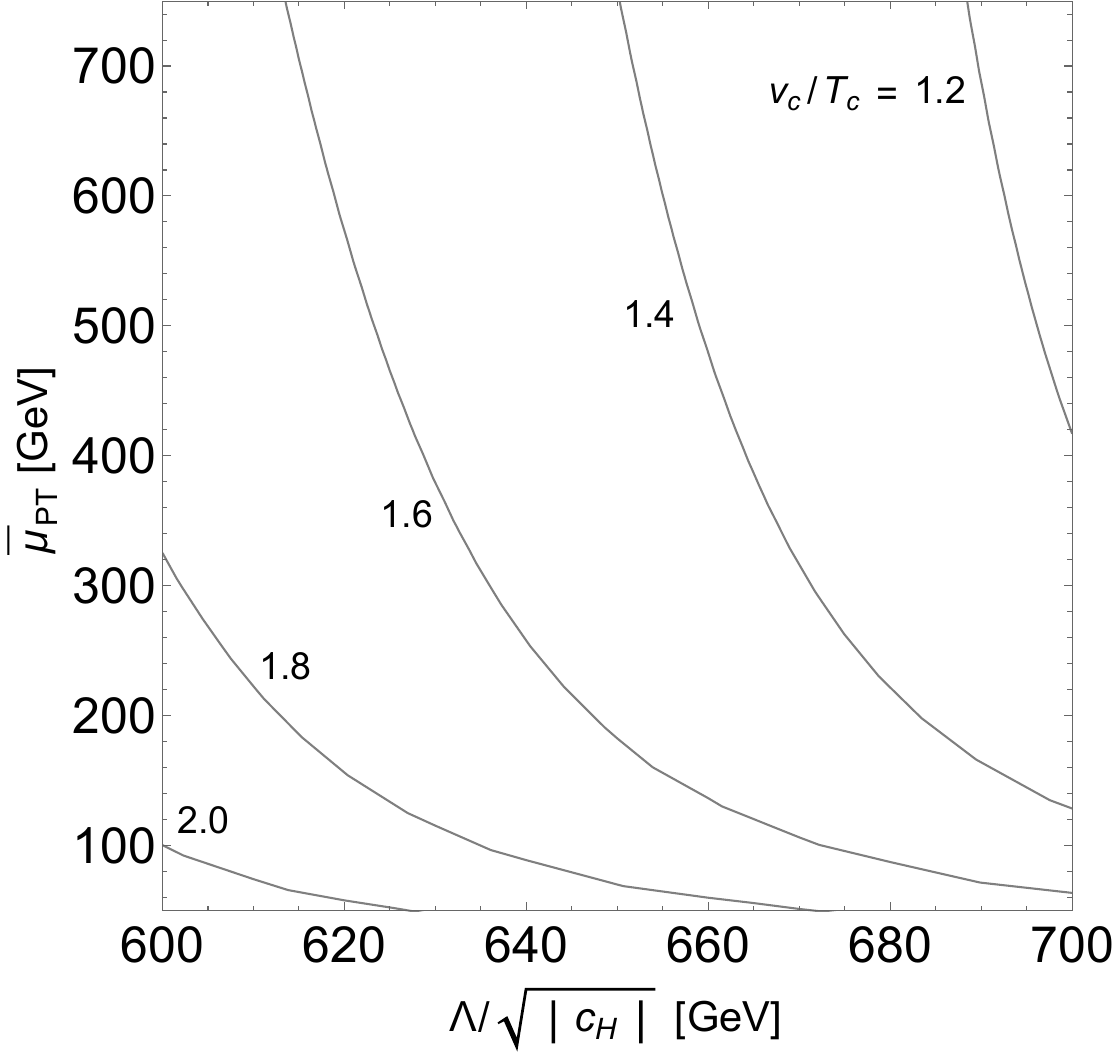}
\end{center}
\vspace{-0.4cm}
\caption{Contour plot of the ratio $v_c/T_c$ as a function of $1/\sqrt{|C_H|} = \Lambda/\sqrt{|c_H|}$ (with $c_H$ dimensionless and $\Lambda$ of mass dimension one), and the renormalization scale $\overline{\mu}_{\rm PT}$, which corresponds to the typical energy scale of the EWPT.
We here consider a scenario where the SFO-EWPT occurs via $\mathcal{O}_H$ and neglect the other operators for brevity.
The other operator effects on the GW spectrum are numerically evaluated in subsection~\ref{sec:sens}.
}
\label{fig:vcTc}
\end{figure}

At the end of this section, we evaluate the ratio $v_c/T_c$, which characterizes the strength of the SFO-EWPT. 
The condition $v_c/T_c \geq 1$ must be satisfied for the SFO-EWPT.
We also assess the associated uncertainty arising from the choice of the renormalization scale.
The critical temperature $T_c$ and the corresponding VEV $v_c$ are defined by the following two conditions:
\begin{align}
    &\partial_{\phi} V_{\rm full}\left(\phi,T_c\right)|_{\phi=v_c}=0,
    \\
    &V_{\rm full}\left(v_c,T_c\right)=V_{\rm full}\left(0,T_c\right),
\end{align}
where $V_{\rm full}\left(\phi,T\right)$, given in eq.~\eqref{eq:Vfull}, is explicitly expressed as a function of $\phi$ and $T$.
In the pure SM, the electroweak phase transition is nonperturbatively known to be a crossover~\cite{Kajantie:1995kf,Kajantie:1996mn,Kajantie:1996qd,Csikor:1998eu,DOnofrio:2015gop}.
However, perturbative calculations fail to predict the endpoint of the line of first-order phase transitions, highlighting the importance of nonperturbative approaches in studying such transitions.
See, {\it e.g.}, Ref.~\cite{Gould:2022ran} for a recent update on a nonperturbative approach.
The impact of the SMEFT operators ($\mathcal{O}_H$, $\mathcal{O}_{H\Box}$, $\mathcal{O}_{HD}$, and $(\mathcal{O}_{uH})_{33}$) on $v_c/T_c$ has been studied in Ref.~\cite{Hashino:2022ghd}, in order to clarify the parameter space of SMEFT where a SFO-EWPT potentially arises.
As it has been shown in Ref.~\cite{Hashino:2022ghd} that the tree-level effects of $\mathcal{O}_H$ dominate the SFO-EWPT, we focus throughout this work on a scenario in which the SFO-EWPT is driven by $\mathcal{O}_H$, while the other SMEFT operators provide small modifications to the full effective potential $V_{\rm full}$.
In Fig.~\ref{fig:vcTc}, we present a numerically calculated contour plot of $v_c/T_c$ as a function of the SMEFT Wilson coefficient $1/\sqrt{|C_H|} := \Lambda/\sqrt{|c_H|}$, where $c_H$ is a dimensionless parameter and $\Lambda$ is a parameter with mass dimension one. 
The plot is also shown as a function of the renormalization scale $\overline{\mu}_{\rm PT}$.
Given that the phase transition parameters depend on $v_c/T_c$, {\it e.g.}, $\alpha \propto (v_c/T_c)^2$~\cite{Espinosa:2010hh}, this result indicates that observables related to the SFO-EWPT, such as the GW spectrum, can be significantly affected by uncertainties in the renormalization scale, as demonstrated in Ref.~\cite{Croon:2020cgk}.

\begin{comment}
\section{First-order phase transition in the SMEFT}
\label{sec:PT}
%
To evaluate the EWPT in the SMEFT, we have to include the thermal potential in addition to the zero temperature potential explained in section~\ref{sec:form}.
%
In subsection~\ref{sec:thermal}, we firstly provide the one-loop thermal effective Higgs potential, including the SMEFT dimension-six operator effects. 
%
Subsequently, in subsection~\ref{sec:proc}, procedures to evaluate the EWPT in this work, including the SMEFT effects, are explained.
%
In subsection~\ref{sec:scale_unc}, we evaluate the ratio $v_c/T_c$ in the SMEFT, quantifying the SFO-EWPT and revisiting the RGE scale uncertainty in the EWPT studied in Ref.~\cite{Croon:2020cgk}.
\end{comment}

\section{RGE effects on GW observations}
\label{sec:GW_sec}
We now examine the RGE effects on the GW spectrum arising from the SFO-EWPT, using the SMEFT as a reference model.
In particular, using a Fisher matrix analysis based on the daisy-resummed approach, we study how precisely future GW observations can measure the SMEFT dimension-six operators listed in Table~\ref{tab:dim6}, assuming that the SFO-EWPT is induced by the $(H^{\dagger}H)^3$ operator and taking into account renormalization scale uncertainties.
In subsections~\ref{sec:GW} and \ref{sec:sta}, we briefly review the evaluation of the GW spectrum and the statistical analysis of the GW signal to facilitate the subsequent discussion.
In subsection~\ref{sec:sens}, we present the sensitivity of future GW observations to the SMEFT operators, taking into account the renormalization scale uncertainties.
See Figs.~\ref{fig:contour}, \ref{fig:DECIGO05}, \ref{fig:DECIGO02}, and \ref{fig:DECIGO1} for the main results.

\subsection{GW spectrum from first-order phase transition}
\label{sec:GW}
%
%We use the GW observation experiments to precisely probe the high dimensional operators and evaluate the confidence intervals for the experiments. 
%We briefly introduce how to evaluate spectra of GW from first-order PT and to estimate the interval. 
%
The spectra of GW from the first-order phase transition can be featured by following four phase transition parameters:
\begin{align}
    T_n,\quad \alpha,\quad \beta/H,\quad v_b.
\end{align}
See, {\it e.g}., Ref.~\cite{Athron:2023xlk} for a review of cosmological phase transitions and gravitational waves.
The first parameter $T_n$ corresponds to the nucleation temperature, at which one bubble is nucleated per horizon on average. 
This temperature is defined as
\begin{align}
 \left.   {\Gamma}/{H^4}\right|_{T=T_n}=1,
\end{align}
where $H$ is the Hubble parameter, and $\Gamma$ is a bubble nucleation rate per unit time and per unit volume,
\begin{align}
	\Gamma\simeq T^4\left(\frac{S_3}{2\pi T} \right)^{3/2} \exp (-S_3/T),\label{eq:Gam}
\end{align}
with a $O(3)$ symmetric bounce solution $S_3$. 
In our numerical analysis, we use the {\tt AnyBubble}~\cite{Masoumi:2016wot} package to calculate the bounce solution.

The second parameter $\alpha$ is the normalized latent heat released by the first-order phase transition and is defined as follows:
%%%%%%%%%%%%%%%%%%%%%%%%%%%%%%%%%%%%%%%%%%
 \begin{align}
 \label{latenth}
 	\alpha:= \epsilon(T_n)/ \rho_{\rm rad}(T_n),
\end{align}
%%%%%%%%%%%%%%%%%%%%%%%%%%%%%%%%%%%%%%%%%%
where $ \epsilon(T)
  =  \Delta V_{\rm full} -T
  \frac{\partial  \Delta V_{\rm full} }{\partial T}$, $\Delta V_{\rm full} =  V_{\rm full}(\phi_-(T),T) - V_{\rm full}(\phi_+(T),T)$, and $\rho_{\rm rad}(T)=(\pi^2/30)g_* T^4$ with degrees of freedom in the plasma $g_{\ast}=106.75$. 
  The $\phi_+$ and $\phi_-$ are the order parameters for the broken and unbroken phases, respectively.

\begin{figure}[t]
\begin{center}
\includegraphics[width=9.cm]{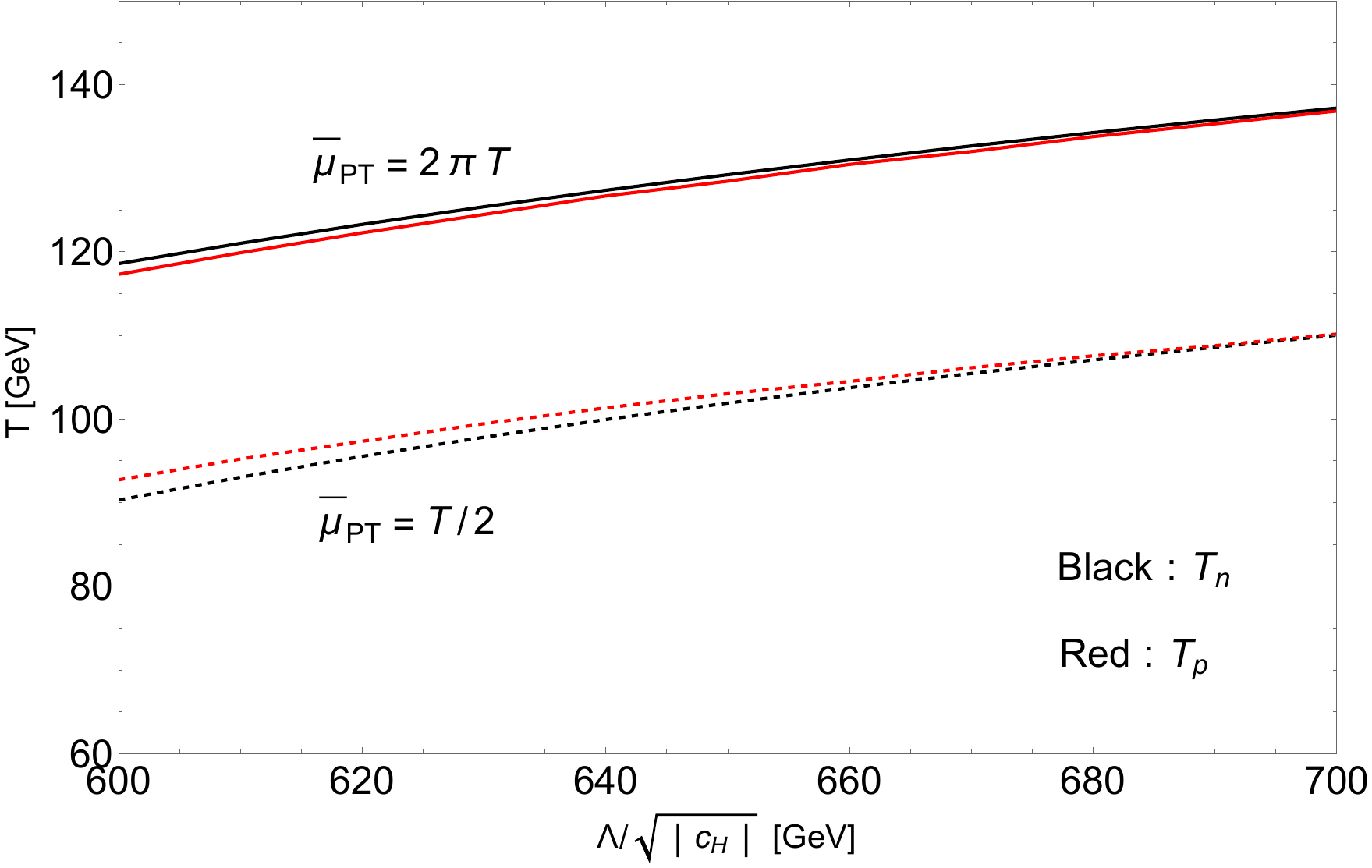}
\end{center}
\vspace{-0.4cm}
\caption{The nucleation temperature $T_n$ (black) and the percolation temperature $T_p$ (red) as functions of $1/\sqrt{|C_H|}=\Lambda/\sqrt{|c_H|}$, where $c_H$ is a dimensionless parameter and $\Lambda$ is a mass-dimensional parameter.
For the solid and dashed curves, we choose different renormalization scales:  $\overline{\mu}_{\rm PT}=2\pi T$ for the solid curve and $\overline{\mu}_{\rm PT}=T/2$ for the dashed curve.
In this plot, all SMEFT operators except $\mathcal{O}_H$ are set to zero to highlight the renormalization scale uncertainty in the potential GW source, {\it i.e.}, $\mathcal{O}_H$.
}
\label{fig:TnTp}
\end{figure}

The third parameter $\beta/H$ corresponds to the inverse of the duration of the first-order phase transition and is defined as
%%%%%%%
	\begin{align} 
	\frac{\beta}{H}:= T_n\left.\frac{d}{dT}\left(\frac{S_3}{T}\right)\right|_{T=T_n}.
	\end{align}
%%%%%%%	
%
For instance, in Refs.~\cite{Ellis:2018mja,Caprini:2019egz,Croon:2020cgk,Bernardo:2025vkz}, these phase transition parameters are evaluated at not the nucleation temperature $T_n$ but the percolation temperature $T_p$ (approximately defined by $I(T_p)=0.34$~\cite{Ellis:2018mja}, where $I(t)=\left(\frac{4\pi}{3}\right)\int_{t_c}^tdt' \Gamma(t')a(t')^3r (t,t')^3$ with the Friedmann-Robertson-Walker scale factor $a(t)$ and $r (t,t')=\int_{t'}^t \left(v_b/a(\tilde{t})\right)d\tilde{t}$). 
In Fig.~\ref{fig:TnTp}, we show the two temperatures, $T_n$ (black) and $T_p$ (red), as functions of $1/\sqrt{|C_H|} = \Lambda/\sqrt{|c_H|}$, where $c_H$ is a dimensionless parameter and $\Lambda$ is a mass-dimensional parameter, for two renormalization scales: $\overline{\mu}_{\rm PT} = 2\pi T$ (solid) and $T/2$ (dashed).
When focusing on the same-colored curves, the difference between the solid and dashed lines quantifies the theoretical uncertainty associated with the renormalization scale $\overline{\mu}_{\rm PT}$.
In contrast, when focusing on a single type of curve, either solid or dashed, the difference between the black and red curves represents the discrepancy between $T_n$ and $T_p$.
Throughout the rest of our analysis, we use the nucleation temperature $T_n$ to evaluate the phase transition parameters.

%%%%%%%%%%%%%%%
%%%%%%%%%%%%%%%
%%%%%%%%%%%%%%%
%%%%%%%%%%%%%%%

The last parameter $v_b$ denotes the bubble wall velocity.
In our analysis, this velocity is treated as a free parameter. 
However, Ref.~\cite{Lewicki:2021pgr} suggests that $v_b$ in the SMEFT scenario can range from approximately 0.2 to 0.8.
Following this, we set 
$v_b$ = 0.2, 0.5, and 1 in our analysis to examine the theoretical uncertainty in the GW spectrum arising from this variation.

%Using these PT parameters, we can evaluate the spectra of GW from first-order PT. 

Next, we will summarize the GW spectrum produced by the first-order phase transition, based on the aforementioned phase transition parameters.
The GW spectrum from the first-order phase transition arises from three sources: plasma compression waves, plasma turbulence, and bubble collisions.
Since the contribution from compression waves dominates the GW spectrum, we focus on this effect in our numerical analysis.
The GW spectrum of compression waves during the phase transition in the radiation era can be represented by the following fitting function~\cite{Caprini:2019egz,Hindmarsh:2017gnf}:
%
%%%%%%%%%%%%%%%
	\begin{align}
  \Omega_{\rm comp} (f) 
  &= 2.061 F_{\rm gw,0}\tilde{\Omega}_{\rm gw} \left(\frac{f}{\tilde{f}_{\rm comp}}\right)^3
  \left(\frac{7}{4+3(f/\tilde{f}_{\rm comp})^2}\right)^{7/2}\notag
  \\
  &\quad\quad\quad\quad\quad\times\begin{cases}
  \left(\frac{\kappa_v\alpha}{1+\alpha}\right)^2  (H_*R_*)
   ,\quad{\rm for}~H_*R_* \leq \sqrt{\frac{3}{4}\kappa_v\alpha/(1+\alpha)}
   \\
    \left(\frac{\kappa_v\alpha}{1+\alpha}\right)^{3/2} (H(T_n)R_*)^2,\quad{\rm for}~ \sqrt{\frac{3}{4}\kappa_v\alpha/(1+\alpha)} < H_*R_* 
  \end{cases}  \label{eq:GWspcom} 
	  \end{align}
%%%%%%%%%%%%%%%
where $\kappa_v$ is an efficiency factor given as~\cite{Espinosa:2010hh},
 %%%%%%%%%%%%%%%
\begin{align}
  \kappa_v(v_b, \alpha)\simeq
  \begin{cases}
    & \frac{ c_s^{11/5}\kappa_A \kappa_B }{(c_s^{11/5} -  v_b^{11/5} )\kappa_B
      +  v_b c_s^{6/5} \kappa_A}\quad \text{for}~ v_b \lesssim c_s  \\
    & \kappa_B + ( v_b - c_s) \delta\kappa 
    + \frac{( v_b - c_s)^3}{ (v_J - c_s)^3} [ \kappa_C - \kappa_B -(v_J - c_s) 
    \delta\kappa ]
    \quad \text{for}~ c_s <  v_b < v_J \\
    & \frac{ (v_J - 1)^3 v_J^{5/2}  v_b^{-5/2}
      \kappa_C \kappa_D }
    {[( v_J -1)^3 - ( v_b-1)^3] v_J^{5/2} \kappa_C
      + ( v_b - 1)^3 \kappa_D }
    \quad \text{for}~ v_J \lesssim v_b 
  \end{cases}.     \label{eq:effi} 
\end{align}
%%%%%%%%%%%%%%%
The $\kappa_A$, $\kappa_B$, $\kappa_C$, and $\kappa_D$ appearing in eq.~(\ref{eq:effi}) are listed as,
%%%%%%%%%%%%%%%
\begin{align}
  \kappa_A 
  &\simeq v_b^{6/5} \frac{6.9 \alpha}{1.36 - 0.037 \sqrt{\alpha} + \alpha},
  \quad
  \kappa_B 
  \simeq \frac{\alpha^{2/5}}{0.017+ (0.997 + \alpha)^{2/5} },
    \nonumber \\
  \kappa_C 
  &\simeq \frac{\sqrt{\alpha}}{0.135 + \sqrt{0.98 + \alpha}},
  \quad
  \kappa_D 
  \simeq \frac{\alpha}{0.73 + 0.083 \sqrt{\alpha} + \alpha},
\end{align}
%%%%%%%%%%%%%%%
%
with the velocity of sound  $c_s=0.577$, and 
\begin{align}
    v_J=\frac{\sqrt{2/3\alpha +\alpha^2}+\sqrt{1/3}}{1+\alpha},\quad  \delta \kappa\simeq -0.9 \ln \frac{\sqrt{\alpha}}{1+\sqrt{\alpha}}.
\end{align}
Furthermore, the remaining relevant quantities are as follows: $F_{\rm gw,0} = 3.57 \times 10^{-5} \left(100/g_\ast\right)^{1/3}$, $\tilde{\Omega}_{\rm gw}=1.2 \times 10^{-2}$, $H_*R_*=(8\pi)^{1/3}(\beta/H)^{-1} $ max($c_s,v_b$), and the peak frequency $\tilde{f}_{\rm comp}$ given by,
%
%%%%%%%%%%%%%%%
	\begin{align}
  \tilde{f}_{\rm comp} \simeq 26 \left(\frac{1}{H_*R_*}\right) 
  \left(\frac{T_n}{100~{\rm GeV}}\right)
  \left(\frac{g_\ast}{100}\right)^{1/6} ~\mu{\rm Hz}.
	\end{align}
%%%%%%%%%%%%%%%
%
The SMEFT effects appearing in the effective potential~\eqref{eq:Vfull} are reflected in the phase transition parameters mentioned above, potentially influencing GW observations.

%The GW spectra from plasma turblance is given by ....

\subsection{Statistical analysis in GW experiments}
\label{sec:sta}
Now, we briefly summarize the statistical analysis used to quantify how precisely future GW observations can measure the SMEFT Wilson coefficients.
As discussed in Ref.~\cite{Hashino:2022ghd}, the uncertainty in the Wilson coefficients indicates a potential window for NP effects when their central values are consistent with zero within the uncertainty.
To estimate the confidence intervals for the Wilson coefficients based on GW observations, we consider the logarithm of the likelihood function, which is approximated as follows~\cite{Seto:2005qy,Hashino:2018wee,Hashino:2022ghd}:
%
%%%%%%%%%%%%%%%
\begin{align}
  \delta \chi^2(\{p\},\{\hat{p}\})\simeq {\cal F}_{ab}(p_a-\hat{p_a})(p_b-\hat{p_b}),\label{eq:chi}
\end{align}
%%%%%%%%%%%%%%%
where ${p}$ denotes the set of Wilson coefficient parameters in the SMEFT, ${\hat{p}}$ represents their fiducial values, and ${\cal F}_{ab}$ is defined as the Fisher information matrix~\cite{Seto:2005qy},
%%%%%%%%%%%%%%%
\begin{align}
{\mathcal F}_{ab}
&=
2T_{\rm obs}
\int_0^\infty df
~
\frac{\partial_{p_a} S_h(f,\left\{ \hat{p} \right\}) \partial_{p_b} S_h(f,\left\{ \hat{p} \right\})}
{\left[ S_{\rm eff}(f) + S_h(f,\left\{ \hat{p} \right\}) \right]^2}.
\label{eq:FabSeff}
\end{align}
%%%%%%%%%%%%%%%
%
Here, $T_{\rm obs}$ represents the observation period in the experiment, while $S_h$ denotes the power spectrum, which is defined as follows:
%
%%%%%%%%%%%%%%%
\begin{align}
S_h(f)
&= 
\frac{3H_0^2}{2\pi^2}
\frac{1}{f^3}
\Omega_{\rm GW}(f),\label{eq:pow}
\end{align}
%%%%%%%%%%%%%%%
%
where $\Omega_{\rm GW}\simeq \Omega_{\rm comp}$.
In the following analysis of GW observations, we focus on a scenario in which only two SMEFT operators (the GW source $\mathcal{O}_H$ and another operator in Table~\ref{tab:dim6}) are active while all others are set to zero.
Then, we derive the confidence intervals on two-dimensional planes spanned by the two SMEFT Wilson coefficients associated with the parameter set $\{p\}=\{1/\sqrt{|C_H|},\,C_{\rm SMEFT}\}$, where $C_{\rm SMEFT}$ denotes one of the SMEFT Wilson coefficients appearing in Table~\ref{tab:dim6}, excluding $C_H$.
The 95\% C.L. interval for the Wilson coefficients is given by $\delta \chi^2 = 6.0$ on the two-dimensional planes.
Based on Refs.~\cite{Yagi:2011wg,Klein:2015hvg}, we present the effective sensitivities of the experiments LISA, DECIGO, and BBO as follows:
%%%%%%%%%%%%%%%
\begin{itemize}
\item LISA

\begin{align}
    S_{\rm eff}(f)=\frac{20}{3}\frac{4 S_{\rm acc}(f)+S_{\rm sn}(f)+S_{\rm omn}(f)}{L^2}\left[1+\left(\frac{f}{0.41 c/2L}\right)^2\right],
\end{align}
with $L=5\times 10^9$~m and,
\begin{align}
    &S_{\rm acc}(f)=9\times 10^{-30} \frac{1}{(2\pi f/1{\rm Hz})^4}\left(1+\frac{10^{-4}}{f/1{\rm Hz}}\right)~{\rm m^2 Hz^{-1}},
    \\
    &S_{\rm sn}(f)=2.96\times 10^{-23}~{\rm m^2 Hz^{-1}},
    \\
    &S_{\rm omn}(f)=2.65\times 10^{-23}~{\rm m^2 Hz^{-1}}.
\end{align}

\item
DECIGO
\begin{align}
S_{\rm eff}(f) 
&=
\left[
\frac{}{}
7.05 \times 10^{-48} 
\left[
1 + (f / f_p)^2
\right]
\right.
\nonumber \\
&~~~~
\left.
+ 
4.8 \times 10^{-51} 
\frac{(f / 1{\rm Hz})^{-4}}{1 + (f / f_p)^2} 
+
5.33 \times 10^{-52}
(f / 1{\rm Hz})^{-4}
\right]
~{\rm Hz^{-1}},
\label{eq:SeffDECIGO}
\end{align}
with $f_p = 7.36$~Hz.
\item
BBO 
\begin{align}
S_{\rm eff}(f)
&=
\left[
2.00 \times 10^{-49} 
(f / 1{\rm Hz})^2
+ 
4.58 \times 10^{-49}
+
1.26 \times 10^{-52}
(f / 1{\rm Hz})^{-4}
\right]
~{\rm Hz^{-1}}.
\label{eq:SeffBBO}
\end{align}
\end{itemize}
%%%%%%%%%%%%%%%

Stochastic GWs from astrophysical sources are known to contribute as a foreground.
In our numerical calculations, we include the foreground from compact white dwarf binaries in our Galaxy, which contributes in the milli-Hertz regime, into the effective sensitivity of each experiment.
The noise spectrum from white dwarfs is evaluated as follows~\cite{Klein:2015hvg}:
%
%%%%%%%%%%%%%%%
\begin{align}
    S_{\rm WD}(f)=\frac{1}{1/S_{\rm WD}^{(1)}(f)+1/S_{\rm WD}^{(2)}(f)+1/S_{\rm WD}^{(3)}(f)+1/S_{\rm WD}^{(4)}(f)},
\end{align}
%%%%%%%%%%%%%%%
where 
%%%%%%%%%%%%%%%
\begin{align}
S'_{\rm WD}(f)=
\begin{cases}
(20/3) (f / 1~{\rm Hz})^{-2.3} \times 10^{-44.62}~{\rm Hz^{-1}}&\equiv S^{(1)}_{\rm WD}(f)
\quad(10^{-5}~{\rm Hz} < f < 10^{-3}~{\rm Hz}), \\
(20/3) (f / 1~{\rm Hz})^{-4.4} \times 10^{-50.92}~{\rm Hz^{-1}}&\equiv S^{(2)}_{\rm WD}(f)
\quad (10^{-3}~{\rm Hz} < f < 10^{-2.7}~{\rm Hz}), \\
(20/3) (f / 1~{\rm Hz})^{-8.8} \times 10^{-62.8}~{\rm Hz^{-1}}&\equiv S^{(3)}_{\rm WD}(f)
\quad (10^{-2.7}~{\rm Hz} < f < 10^{-2.4}~{\rm Hz}), \\
(20/3) (f / 1~{\rm Hz})^{-20.0} \times 10^{-89.68}~{\rm Hz^{-1}}&\equiv S^{(4)}_{\rm WD}(f)
\quad(10^{-2.4}~{\rm Hz} < f < 10^{-2}~{\rm Hz}). 
 \end{cases}
\end{align}
%%%%%%%%%%%%%%%
%
In Fig.~\ref{fig:GW}, we present the sensitivity curves for LISA (blue), DECIGO (green), and BBO (orange).
Additionally, the noise spectrum from the white dwarf is depicted as an orange-red curve.

\subsection{Sensitivity of GW observation to NP effects}
\label{sec:sens}

\begin{figure}[t]
\begin{center}
\includegraphics[width=13.cm]{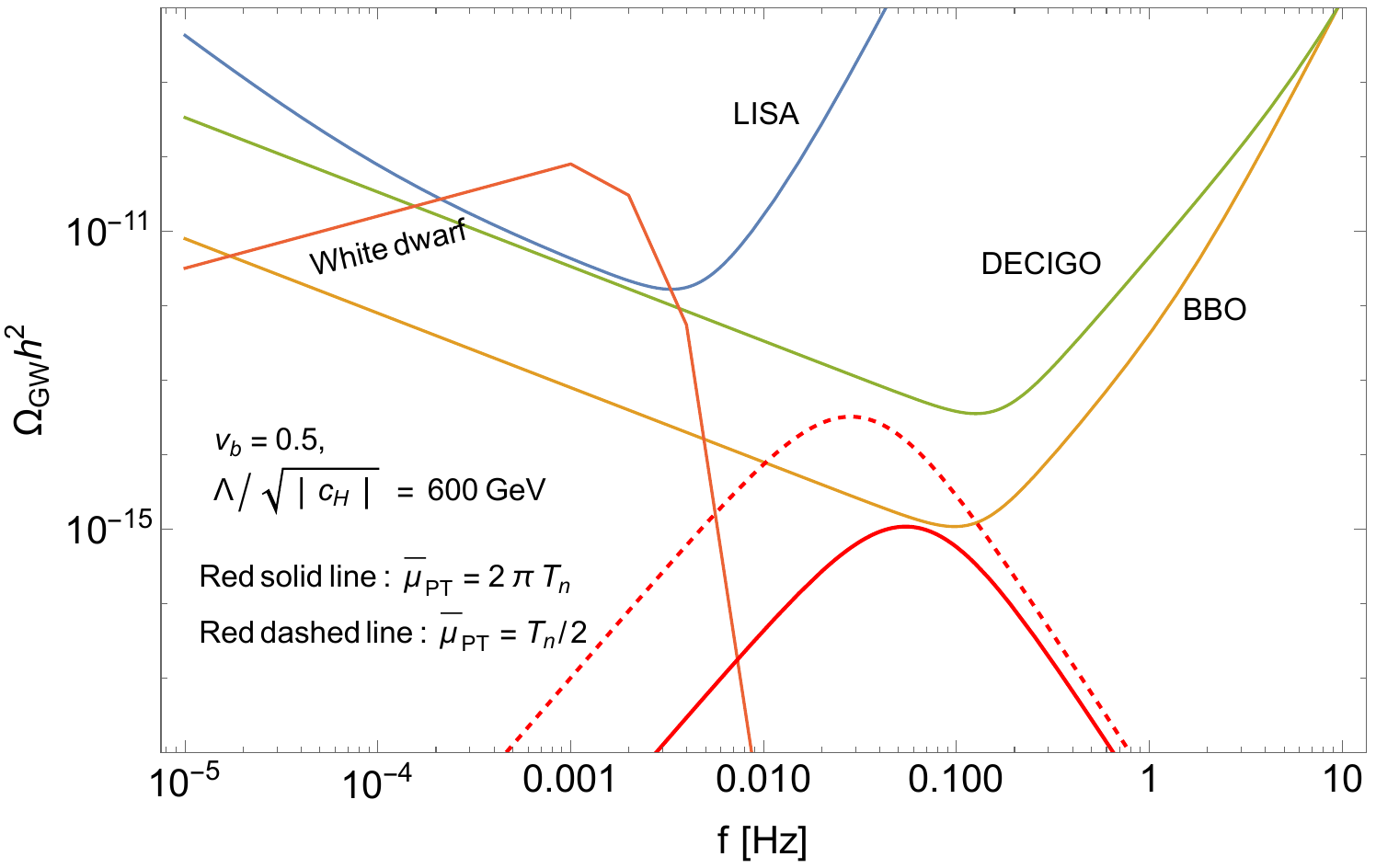}
\end{center}
\vspace{-0.4cm}
\caption{
The dependence of the renormalization scale $\overline{\mu}_{\rm PT}$ in the GW spectra from the SFO-EWPT induced by the SMEFT operator $\mathcal{O}_H$.
The red curves correspond to $\overline{\mu}_{\rm PT}=2\pi T_n$ (solid) and $T_n/2$ (dashed), respectively, under the assumption that $1/\sqrt{|C_H|}=\Lambda/\sqrt{|c_H|}=600$ GeV and $v_b=0.5$.
The blue, green, and orange curves represent the sensitivity of LISA, DECIGO, and BBO, respectively.
The orange-red curve corresponds to the foreground originating from compact white dwarf binaries in our Galaxy.
}.
\label{fig:GW}
\end{figure}

We now provide the results from the numerical analysis of the GW spectrum and its sensitivity to the SMEFT operators.
In Fig.~\ref{fig:GW}, we present the GW spectra from the SFO-EWPT, generated via the SMEFT operator $\mathcal{O}_H$ (red solid and dashed curves). 
In this plot, all SMEFT operators except $\mathcal{O}_H$ are set to zero.
For these two red curves, we select $v_b=0.5$ and $1/\sqrt{|C_H|}=\Lambda/\sqrt{|c_H|}=600$ GeV, where $c_H$ is a dimensionless parameter and $\Lambda$ is a mass-dimensional parameter.  
The renormalization scale is $\overline{\mu}_{\rm PT}=2\pi T_n$ for the solid curve, whereas $\overline{\mu}_{\rm PT}= T_n/2$ corresponds to the dashed curve. 
As discussed in Ref.~\cite{Croon:2020cgk}, the GW spectrum is potentially sensitive to the choice of the renormalization scale $\overline{\mu}_{\rm PT}$, and Fig.~\ref{fig:GW} also illustrates the significant renormalization scale uncertainty.
These results suggest that GW observations cannot precisely determine the central value of the Wilson coefficient for $\mathcal{O}_H$ when applying the daisy-resummation approach, as described in Ref.~\cite {Croon:2020cgk} and Section~\ref{sec:form}.
However, future collider experiments may constrain the central value of the operator $\mathcal{O}_H$~\cite{deBlas:2019rxi}, which is the source of the GW signal in this work.
Once $\mathcal{O}_H$ is determined, it will be necessary to clarify how precisely the GW observations can determine the other SMEFT operators, other than $\mathcal{O}_H$.
According to these, in what follows, we suppose that the Wilson coefficient $C_H$ is determined, {\it e.g.,} by future collider experiments, and perform the Fisher matrix analysis.
It should be noted that we also vary the central value of $C_H$ to study its dependence, as will be detailed later.

%%%%%%%%%%%%%%%
%%%%%%%%%%%%%%%
%%%%%%%%%%%%%%%
%%%%%%%%%%%%%%%

\begin{figure}[t]
\begin{center}
\includegraphics[width=10.cm]{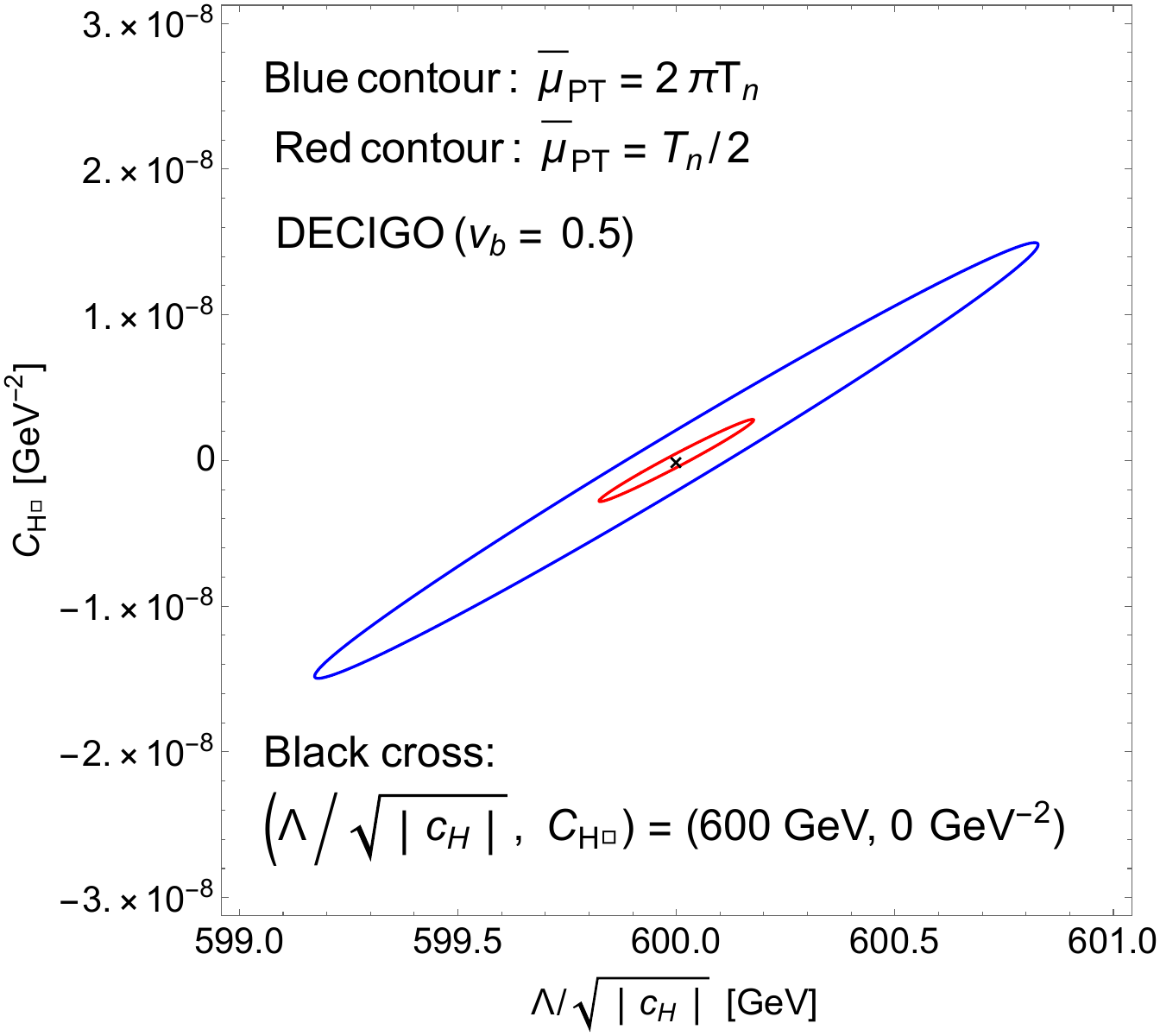}
\end{center}
\vspace{-0.4cm}
\caption{
The 95\% C.L. contours of DECIGO observation with 1-year statistics for $\overline{\mu}_{\rm PT}=2\pi T_n$ (blue) and $T_n/2$ (red), assuming the central value (black cross) of $1/\sqrt{|C_H|}=\Lambda/\sqrt{|c_H|}=600$ GeV and $C_{H\square}=0~{\rm GeV}^{-2}$.
}
\label{fig:contour}
\end{figure}

From eqs.~\eqref{eq:chi} and \eqref{eq:FabSeff}, we evaluate the 95\% confidence intervals, {\it i.e.} $\delta\chi^2=6.0$, in the DECIGO and BBO experiments, assuming a 1-year observation time ($T_{\rm obs} = 1$ year).
By varying the central value of $C_H$, {\it i.e.}, the GW source, we analyze the confidence intervals on the two-dimensional planes defined by the parameter set $\{p\}=\{\Lambda/\sqrt{|c_H|},\,C_{\rm SMEFT}\}$, where $C_{\rm SMEFT}$ represents one of the SMEFT Wilson coefficients listed in Table~\ref{tab:dim6} except for $C_H=c_H/\Lambda^2$.
Unlike $C_H$, whose central value is varied, we fix the central values of the other SMEFT operators $C_{\rm SMEFT}$ to zero. This choice is made to highlight the potential sensitivity of GW observations to the NP scale associated with $C_{\rm SMEFT} := c_{\rm SMEFT}/\Lambda^2$, where $c_{\rm SMEFT}$ is a dimensionless coefficient and $\Lambda$ is a parameter with mass dimension one.
In Fig.~\ref{fig:contour}, we show the 95\% confidence intervals (blue and red curves) for the DECIGO experiment on $\{\Lambda/\sqrt{|c_H|},\,C_{H\square}\}$ plane as a typical result of this analysis.
Here, the central point (black cross) is set to $\{\Lambda/\sqrt{|c_H|},\,C_{H\square}\}=\{600~{\rm GeV},\,0~{\rm GeV}^{-2}\}$, and the wall velocity is set to $v_b=0.5$. 
For the blue and red curves, the renormalization scale is set to $\overline{\mu}_{\rm PT}=2\pi T_n$ and $T_n/2$, respectively.
For instance, the width of the ellipse along the vertical axis quantifies the precision with which the Wilson coefficient $C_{H\square}$ can be measured.
This plot shows that the SMEFT operator, {\it e.g.}, $C_{H\square}$ is potentially constrained by GW observations, considering the renormalization scale uncertainty, provided that the central value of $C_H$ is determined.
In what follows, we comprehensively apply this analysis to all SMEFT operators in Table~\ref{tab:dim6}, varying the renormalization scale as $\overline{\mu}_{\rm PT}=2\pi T_n$ and $T_n/2$, the wall velocity $v_b=0.2$, 0.5, and 1, and the central value of $C_H$.
Note that the central value of $C_H$ corresponds to the point marked by the black cross in Fig.~\ref{fig:contour}, and the analysis is also performed for different choices of this value, as will be detailed below.

%%%%%%%%%%%%%%%
%%%%%%%%%%%%%%%
%%%%%%%%%%%%%%%
%%%%%%%%%%%%%%%

\begin{figure}[htbp]
\begin{center}
\includegraphics[width=16.cm]{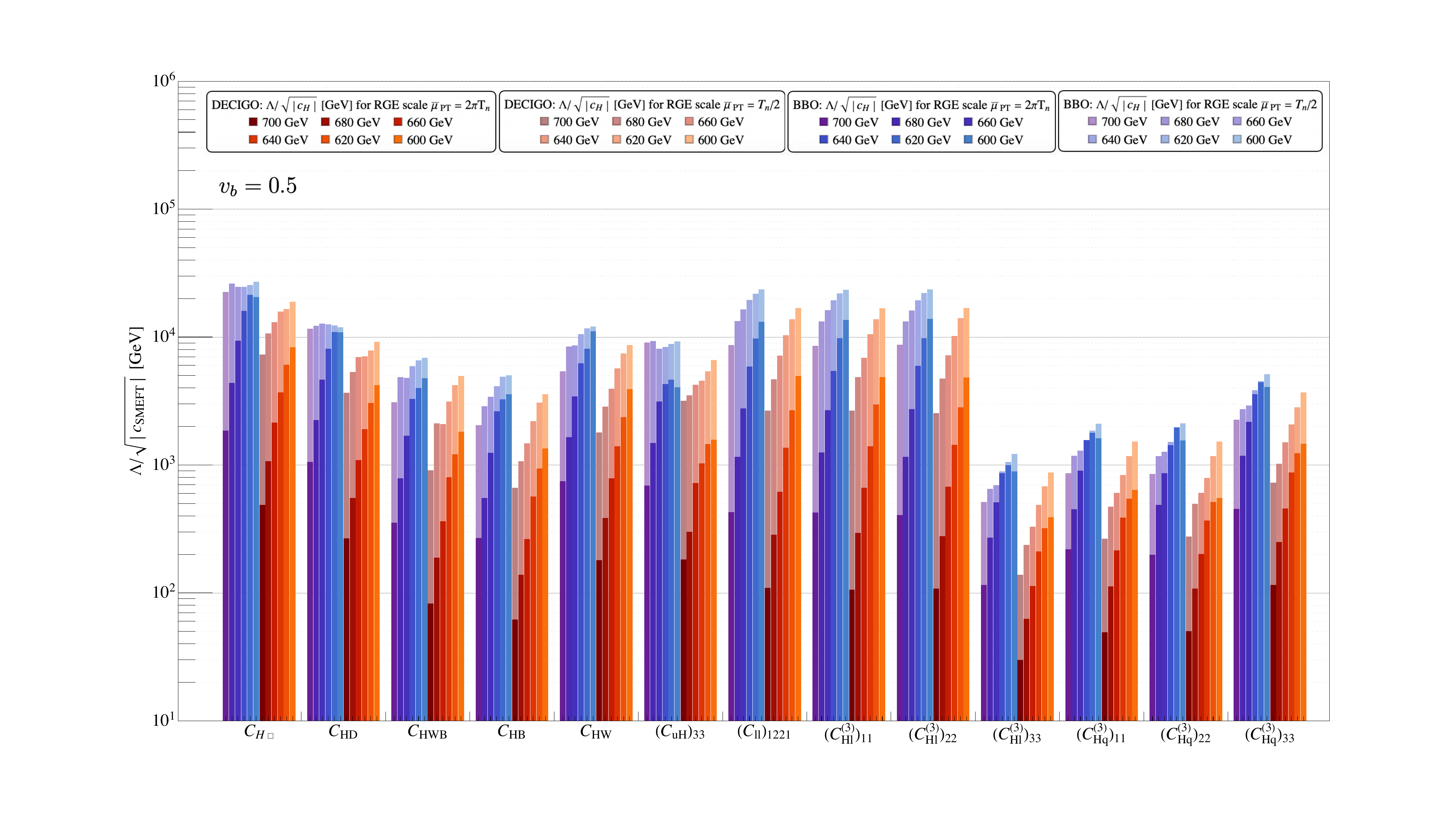}
\includegraphics[width=16.cm]{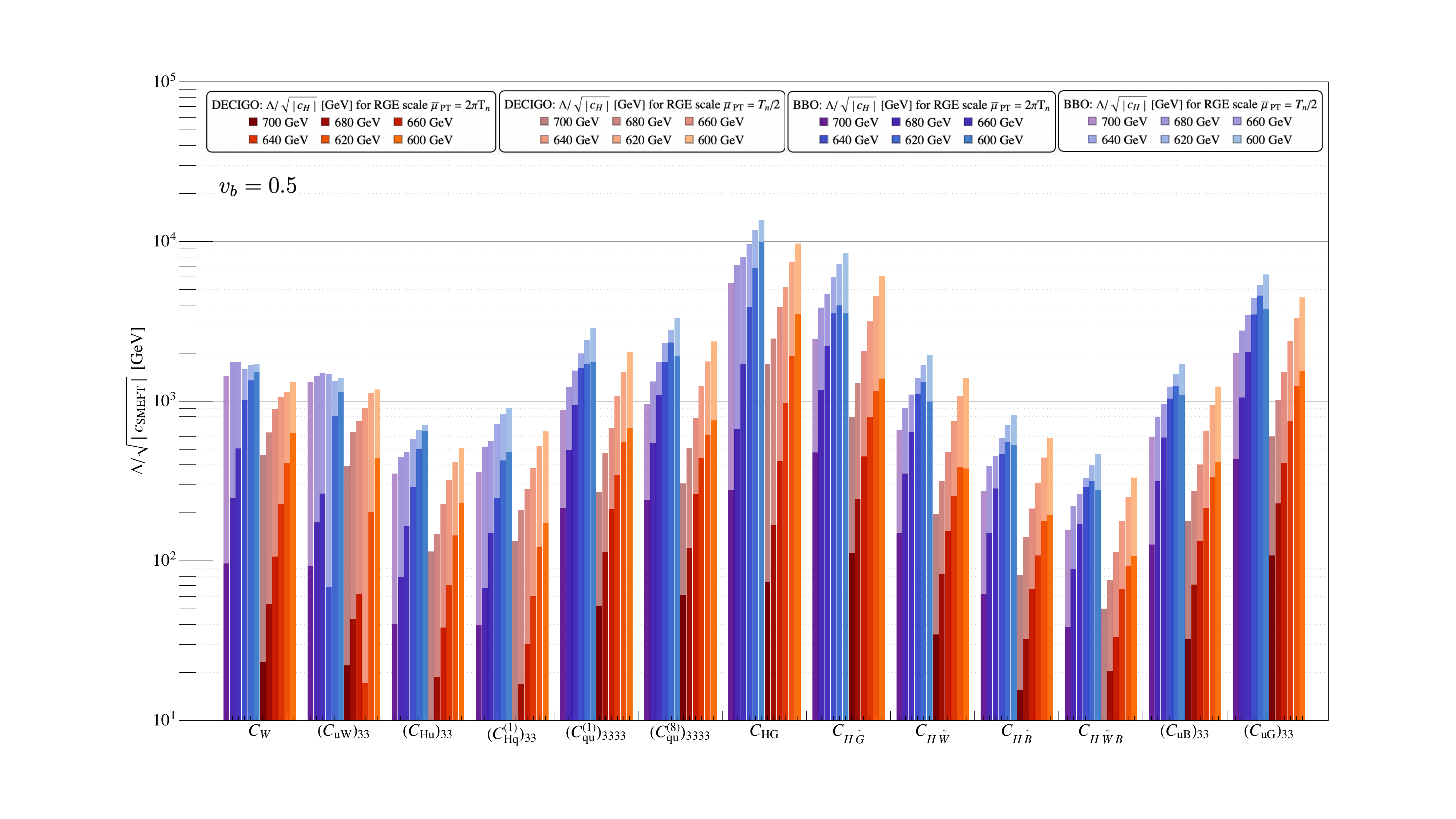}
\end{center}
\vspace{-0.4cm}
\caption{
The projected sensitivity of the DECIGO (reddish bands) and BBO (bluish bands) experiments to $1/\sqrt{|C_{\rm SMEFT}|} := \Lambda/\sqrt{|c_{\rm SMEFT}|}$ (in GeV), where $C_{\rm SMEFT}$ denotes one of the SMEFT Wilson coefficients listed in Table~\ref{tab:dim6}, excluding $C_H$, is shown as a function of the central value of $1/\sqrt{|C_H|} = \Lambda/\sqrt{|c_H|}$.
The vertical axes of these panels represent the width of the 95\% confidence interval for $\Lambda / \sqrt{|c_{\rm SMEFT}|}$, rather than its central value.
The analysis assumes $v_b = 0.5$, a central value of $C_{\rm SMEFT}=c_{\rm SMEFT}/\Lambda^2$ set to $0~\text{GeV}^{-2}$, and an observation time of one year.
For both the dark and light bands, different colors correspond to different central values of $C_H$, {\it i.e.}, $\Lambda/\sqrt{|c_H|} = 600$, 620, 640, 660, 680, and 700 GeV.
See the legend for details.
}
\label{fig:DECIGO05}
\end{figure}

\begin{figure}[htbp]
\begin{center}
\includegraphics[width=16.cm]{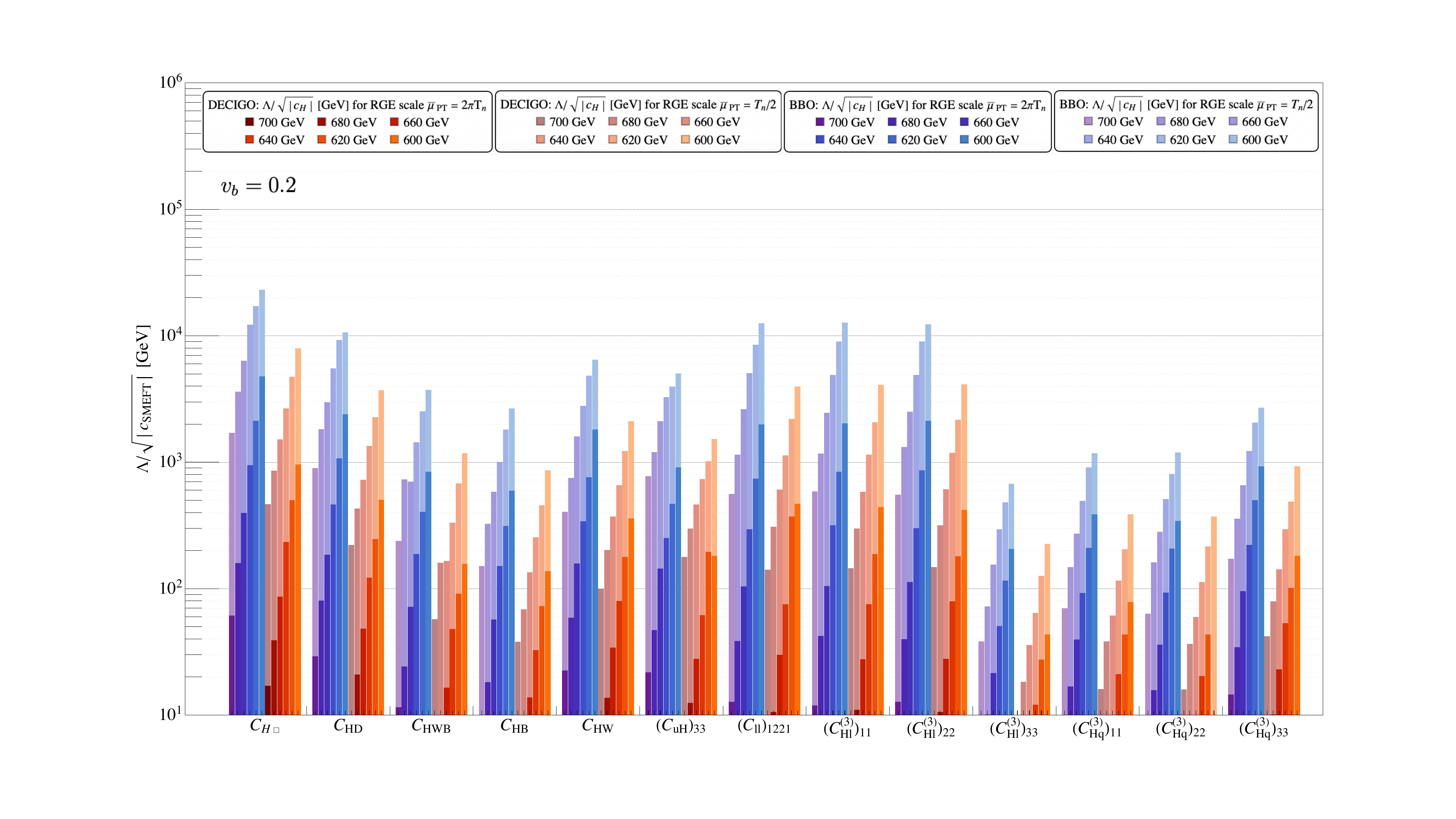}
\includegraphics[width=16.cm]{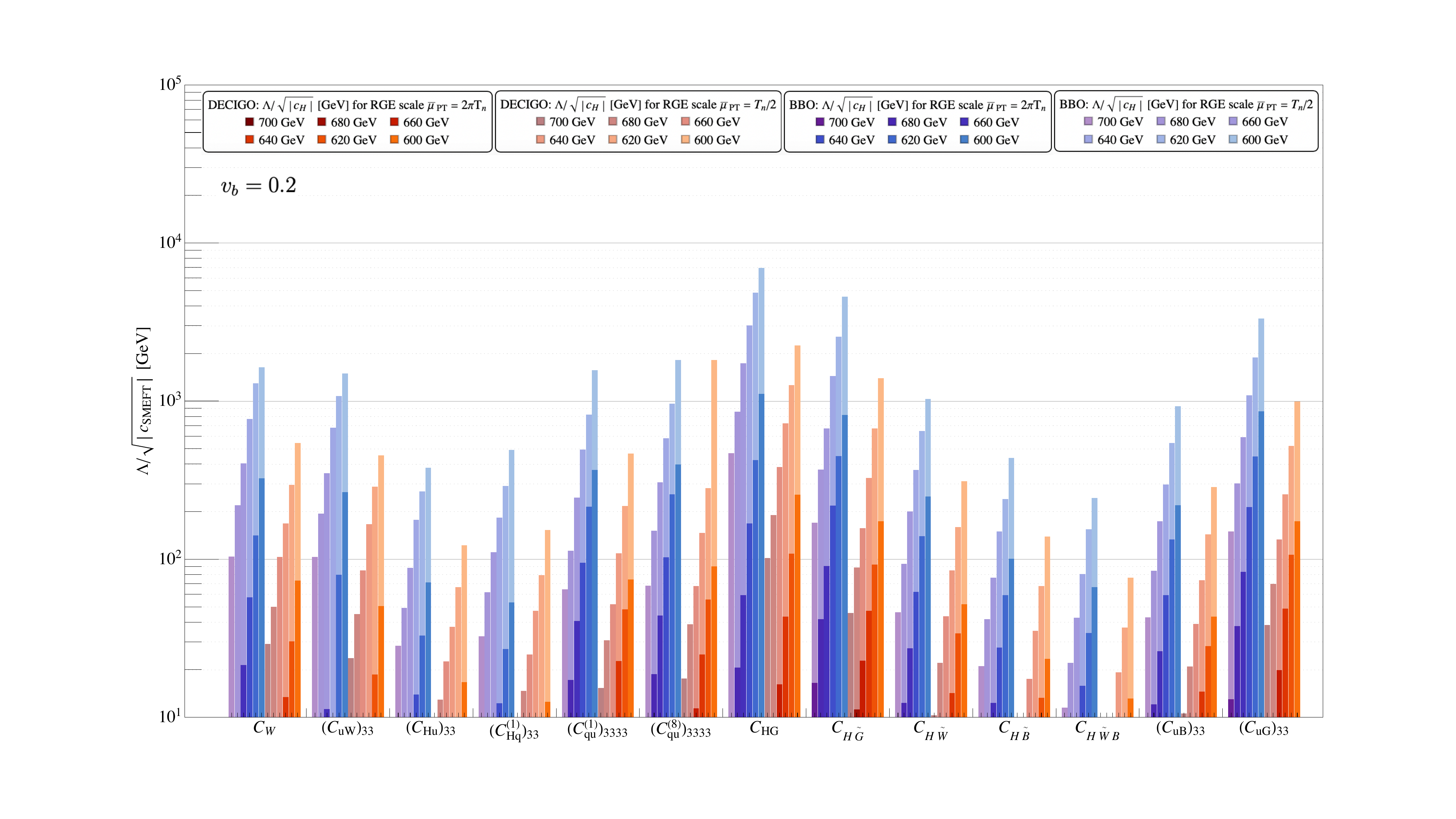}\end{center}
\vspace{-0.4cm}
\caption{The same plots as in Fig.~\ref{fig:DECIGO05}, but for $v_b = 0.2$.
}.
\label{fig:DECIGO02}
\end{figure}

\begin{figure}[htbp]
\begin{center}
\includegraphics[width=16.cm]{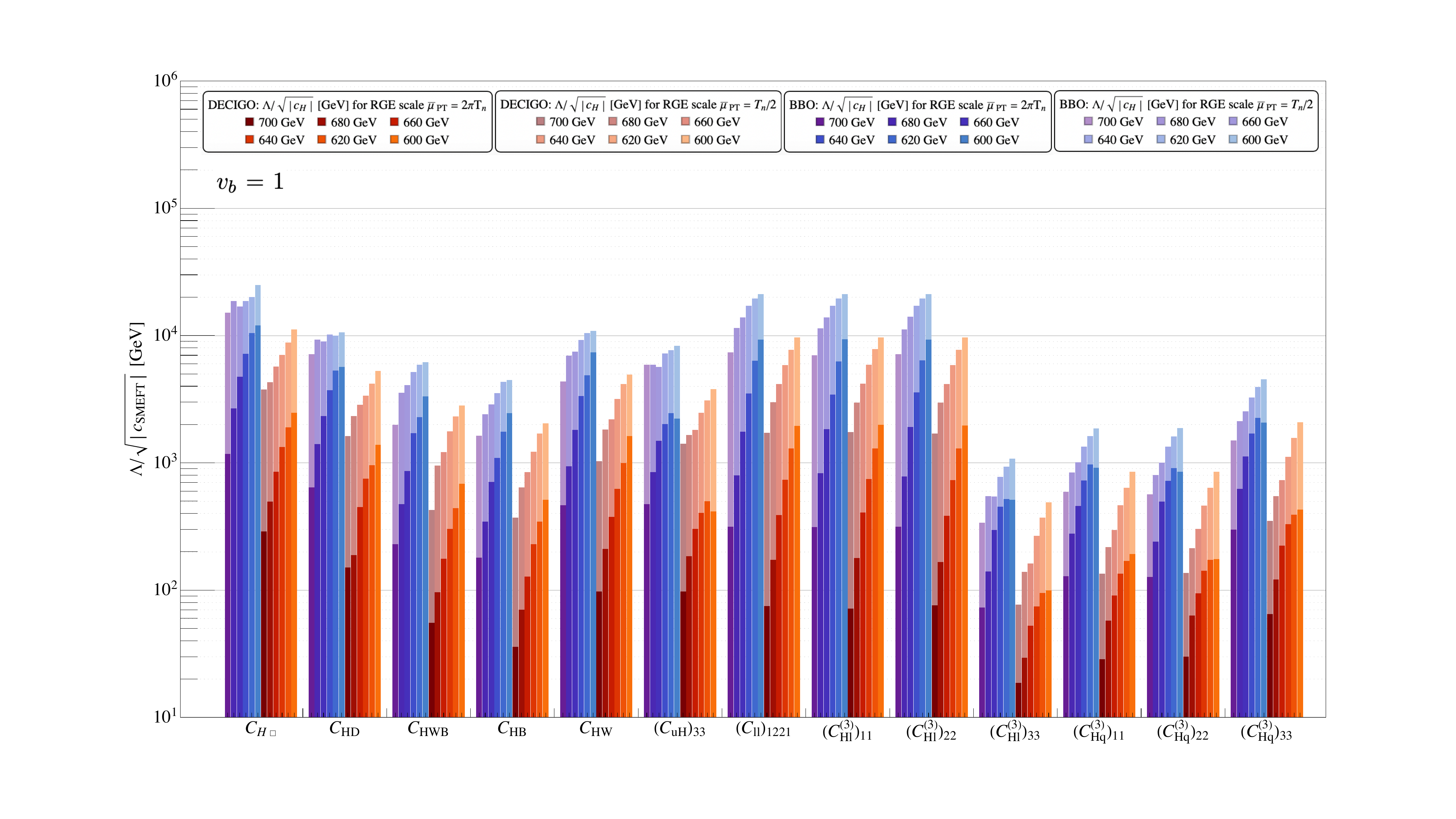}
\includegraphics[width=16.cm]{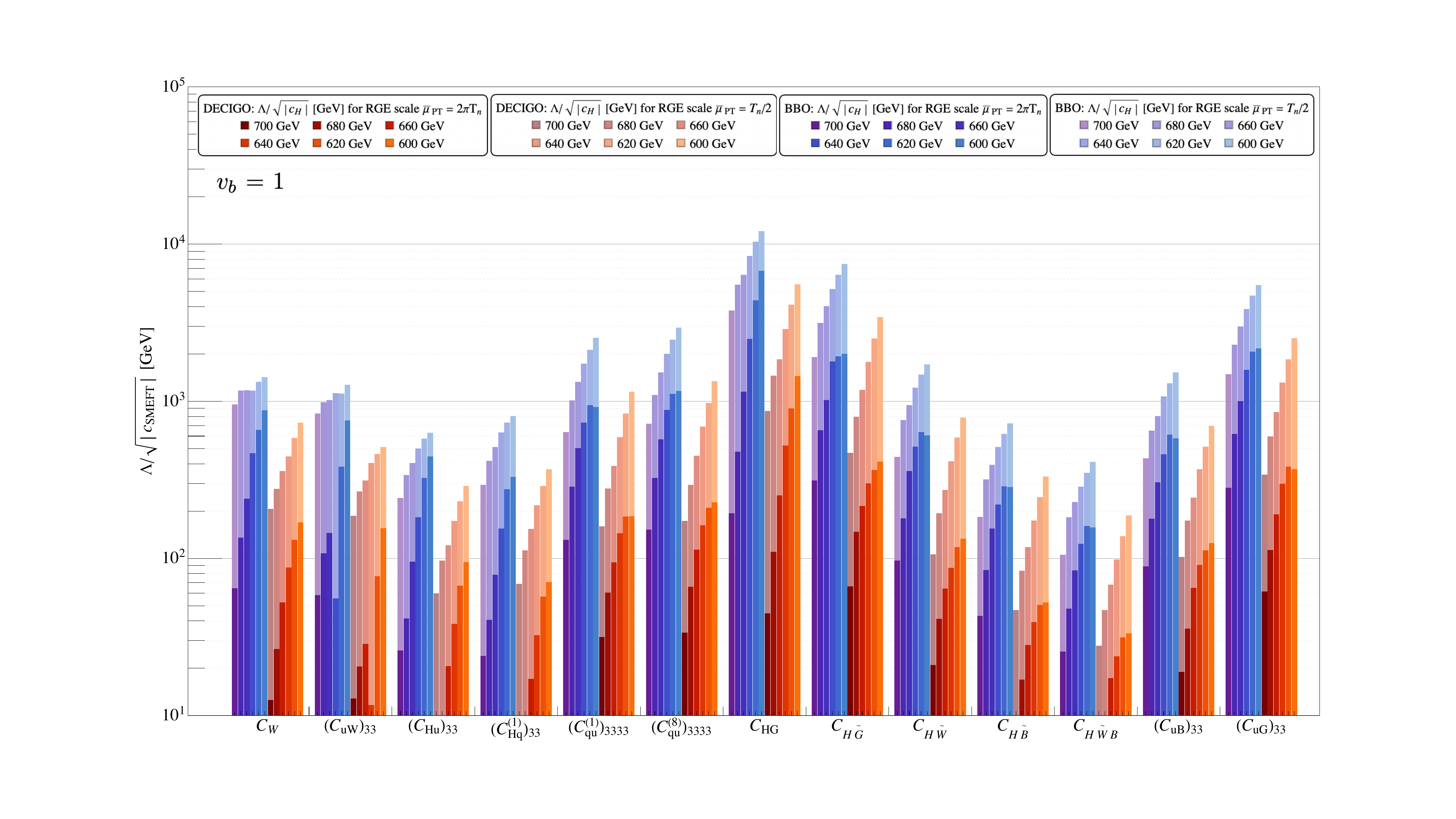}\end{center}
\vspace{-0.4cm}
\caption{The same plots as in Fig.~\ref{fig:DECIGO05}, but for $v_b = 1$.
}.
\label{fig:DECIGO1}
\end{figure}

In Fig.~\ref{fig:DECIGO05}, we present the 95\% confidence level (C.L.) results for DECIGO (reddish bands) and BBO (bluish bands), assuming $v_b = 0.5$ and 1-year observation time.
The 95\% C.L. intervals indicated by the vertical extent of the colored bands align with the vertical size of the ellipses in Fig.~\ref{fig:contour}, for instance, for the $C_{H\Box}$ operator.
The dark- and light-colored bands represent the choices $\overline{\mu}_{\rm PT} = 2\pi T_n$ and $T_n/2$, respectively, reflecting the renormalization scale uncertainty.
Within each shade, different colors indicate different central values of $1/\sqrt{|C_H|} = \Lambda/\sqrt{|c_H|}$, namely, $\Lambda/\sqrt{|c_H|} = 600$, 620, 640, 660, 680, and 700~GeV.
In Ref.~\cite{deBlas:2019rxi}, it was shown that, for example, the 68\% probability reach for $C_H$ at the HL-LHC, ILC1000, and FCC-ee/eh/hh is expected to be $\pm 1.1/(1~\mathrm{TeV})^2$, $\pm 0.2/(1~\mathrm{TeV})^2$, and $\pm 0.1/(1~\mathrm{TeV})^2$, respectively (see, {\it e.g.}, Table 9).
Accordingly, for a set of values $1/\sqrt{|C_H|}$ (in GeV), $\{600, 620, 640, 660, 680, 700\}$, the corresponding uncertainties can be estimated as follows:
\begin{itemize}
    \item HL-LHC: $\{600^{+170}_{-91},\ 620^{+190}_{-99},\ 640^{+220}_{-110},\ 660^{+250}_{-120},\ 680^{+280}_{-130},\ 700^{+320}_{-130}\}$
    \item ILC1000: $\{600^{+22}_{-20},\ 620^{+24}_{-22},\ 640^{+27}_{-24},\ 660^{+30}_{-26},\ 680^{+33}_{-29},\ 700^{+36}_{-31}\}$
    \item FCC-ee/eh/hh: $\{600^{+12}_{-11},\ 620^{+13}_{-12},\ 640^{+14}_{-13},\ 660^{+16}_{-15},\ 680^{+17}_{-16},\ 700^{+19}_{-17}\}$
\end{itemize}
In particular, the uncertainties for ILC1000 and FCC-ee/eh/hh are sufficiently small compared to the overall range of this set, allowing one to visualize their effects by comparing the colored bands in Figs.~\ref{fig:DECIGO05}, \ref{fig:DECIGO02}, and \ref{fig:DECIGO1}.
Although, in general, the uncertainties depend on the central value of $C_H$, they are expected to be independent of it at leading (linear) order in the SMEFT expansion, where the covariance matrix becomes independent of $C_H$.
Further details on the color scheme can be found in the caption of Fig.~\ref{fig:DECIGO05}.
From this figure, we find that future GW observations could accurately constrain the SMEFT Wilson coefficients within renormalization scale uncertainties, provided that other experiments determine the central value of $C_H$.
It is worth emphasizing that the accuracy of future GW observations is well preserved even when the central value of $C_H$, as determined by other experiments, varies from 600 GeV to 700 GeV.
Moreover, our analysis focuses on a scenario in which the SFO-EWPT is driven by the parameter $C_H$.
Therefore, if $C_H$ is treated as a completely free parameter — for example, allowing $C_H = 0$ — gravitational waves are not generated, and the sensitivities shown in Fig.~\ref{fig:DECIGO05} vanish entirely.
Figures~\ref{fig:DECIGO02} and \ref{fig:DECIGO1} show the same plot as Fig.~\ref{fig:DECIGO05}, but with the bubble wall velocity set to $v_b = 0.2$ and $v_b = 1$, respectively.
These results indicate that although the wall velocity introduces a significant source of theoretical uncertainty, future GW observations may still enable precise measurements of the SMEFT Wilson coefficients.

%%%%%%%%%%%%%%%
%%%%%%%%%%%%%%%
%%%%%%%%%%%%%%%
%%%%%%%%%%%%%%%

%%%%%%%%%%%%%%%
%%%%%%%%%%%%%%%
%%%%%%%%%%%%%%%
%%%%%%%%%%%%%%%

%%%%%%%%%%%%%%%
%%%%%%%%%%%%%%%
%%%%%%%%%%%%%%%
%%%%%%%%%%%%%%%

\section{Summary}
\label{sec:summary}
In this paper, we have explored the impact of renormalization scale uncertainties on precision measurements of the Higgs potential via future GW observations, using Fisher matrix analyses based on the daisy-resummed approach.
A previous study~\cite{Croon:2020cgk} focused on the SMEFT as a benchmark model and pointed out significant renormalization scale uncertainties in the GW spectrum, arising from the daisy-resummed method.
On the other hand, other studies~\cite{Hashino:2018wee,Hashino:2022ghd} analyzed the precision with which NP effects can be probed by future GW observations using Fisher matrix analyses based on the daisy-resummed approach, but without incorporating renormalization scale uncertainties.
These studies demonstrated that future GW observations could be competitive with future collider experiments.
This left an open question: can future GW observations retain their precision in probing new physics effects when renormalization scale uncertainties are properly included in the daisy-resummed framework?
To address this question, we have performed Fisher matrix analyses incorporating renormalization scale uncertainties, aiming to evaluate the impact on the measurement of new physics effects through GW signals.
Building upon the previous work on renormalization scale uncertainties in the GW spectrum~\cite{Croon:2020cgk}, we adopted SMEFT benchmark scenarios and considered dimension-six operators that contribute to the Higgs effective potential via one-loop RGE effects, including the Higgs self-coupling, top Yukawa coupling, and gauge couplings, in addition to SMEFT tree-level corrections.

We focused on a scenario in which the $(H^{\dagger}H)^3$ operator induces the SFO-EWPT, while the other dimension-six SMEFT operators listed in Table~\ref{tab:dim6} lead to small shifts in the Higgs potential.
As studied in Ref.~\cite{Croon:2020cgk}, the renormalization scale dependence of the GW peak amplitude calculated using the conventional daisy-resummed approach can vary by one to two orders of magnitude, even when including the RGE running of physical parameters. 
Therefore, it would be challenging to determine the value of the Wilson coefficient $C_H$ solely from GW observations.
Accordingly, we specifically examine the case in which the Wilson coefficient $C_H$ is determined by other experiments, such as future collider measurements~\cite{deBlas:2019rxi}, and numerically evaluate how precisely the remaining dimension-six SMEFT operators listed in Table~\ref{tab:dim6} can be constrained using Fisher matrix analyses.
As shown in Figs.~\ref{fig:contour}, \ref{fig:DECIGO05}, \ref{fig:DECIGO02}, and \ref{fig:DECIGO1}, the precision depends on the central value of the Wilson coefficient $C_H$, the renormalization scale $\overline{\mu}_{\rm PT}$, and the bubble wall velocity $v_b$.
Although these figures indicate that the precision of future GW observations is affected by theoretical uncertainties such as the renormalization scale and the bubble wall velocity, they still show potential sensitivity to new physics scales above 10 TeV.
It is remarkable that the precision of future GW observations remains significant even when such theoretical uncertainties are taken into account.

In discussing future directions, it is important to consider the underlying mechanism responsible for the SFO-EWPT.
The SMEFT $(H^{\dagger}H)^3$ operator qualitatively captures the effects of new physics on the Higgs potential in a model-independent manner and remains a valuable benchmark for examining theoretical uncertainties.
However, it has been pointed out that the SFO-EWPT induced by the SMEFT $(H^{\dagger}H)^3$ operator faces quantitative limitations due to the breakdown of the EFT description.
As a next step, it is essential to explore alternative new physics scenarios beyond the $(H^{\dagger}H)^3$ operator that can realize the SFO-EWPT and to perform similar analyses to evaluate the potential of future GW observations in more realistic settings.
Furthermore, as studied in Ref.~\cite{Croon:2020cgk}, a dimensionally reduced approach has been applied to address the large theoretical uncertainty arising from renormalisation scale dependence.
It has been shown that the dependence of the GW peak amplitude on the renormalisation scale can be reduced by a few orders of magnitude.
Therefore, it is expected that the renormalisation scale uncertainty in the Fisher matrix analysis performed in this paper would also be significantly reduced when the dimensionally reduced approach is applied.
Performing such an analysis based on the dimensionally reduced approach is left as an important future direction.

\section*{Acknowledgements}
DU would like to thank the CERN Theory Department for their financial support and hospitality during this work.
DU is supposed by grants from the ISF (No.~1002/23 and 597/24) and the BSF (No.~2021800).
This work is supported by Grant-in-Aid for Early-Career Scientists 25K17398 (KH) and Grant-in-Aid for Research Activity Start-up 23K19052 (KH).

\appendix

\section{Self energies in the SM}
\label{sec:SM_self}
The self-energy functions in the SM have been calculated in Refs.~\cite{Kajantie:1995dw,Croon:2020cgk} at one-loop level as follows:
\begin{align}
    \Pi_{h,{\rm SM}}(M^2_h)&= \frac{3}{8} \frac{g_{0}^2 M^2_h}{(4\pi)^2}\Bigg[
    -\frac{4}{3}-8 \frac{1}{h^2} -2 h^2 +16 \frac{t^4}{h^2} -\frac{2}{3} z^2 -4 \frac{z^4}{h^2}+3 h^2 F(M_h,M_h,M_h)\notag
    \\
    &+ 4t^2 \left(1-4\frac{t^2}{h^2}\right) F(M_h,M_t,M_t)
    +\frac{2}{3} \frac{h^4- 4h^2 +12}{h^2} F(M_h,M_W,M_W)\notag
    \\
    &+\left(\frac{1}{3}\frac{1}{h^2}-\frac{4}{3}z^2 +4 \frac{z^4}{h^2} \right) F(M_h,M_Z,M_Z)
    -2 h^2 \ln h -8 t^2 \ln t \notag
    \\
    &+ \left(-\frac{2}{3}h^2 +4 z^2\right)\ln z+\left(-4+ 2h^2 +4 t^2-2z^2\right)\ln \left(\frac{\overline{\mu}^2}{M^2_W}\right)
    \Bigg],\label{eq:PihSM}
    \\
    \Pi_{W,{\rm SM}}(M^2_W) &=\frac{3}{8}\frac{g_{0}^2 M_W^2}{(4\pi)^2}\Bigg[
    -\frac{212}{9} -\frac{8}{3} \frac{1}{h^2}-\frac{22}{9} h^2 +\frac{4}{27}\left(
    40 N_f -17
    \right) -\frac{4}{3}t^2 +16 \frac{t^4}{h^2}\notag
    \\
    &+\frac{14}{9}z^2 -\frac{4}{3}\frac{z^4}{h^2}
    +\frac{4 h^2 (h^2-2)}{h^2-1}\ln h-8\left(\frac{2}{3}-t^2 +4 \frac{t^4}{h^2}\right)\ln t \notag
    \\
    &+ 4 \left( 2 \frac{z^4}{h^2}-\frac{z^4-4z^2-8}{z^2-1}\right)\ln z
    +\frac{2}{9}\left(12 -4 h^2 +h^4\right) F(M_W, M_h, M_W)\notag
    \\
    &-\frac{4}{3} (t^2+2)(t^2-1) F(M_W,M_t,0)
    -\frac{32}{3} \frac{z^2-1}{z^2} F(M_W,M_W,0)\notag
    \\
    &
    +\frac{2}{9} \frac{(z^4 + 20 z^2 +12)(z^2-4)}{z^2}F(M_W,M_W,M_Z)\notag
    \\
    &+2 \left\{
    -1+\frac{2}{h^2}+\left(-\frac{59}{9}-6 \frac{1}{h^2}-h^2 +\frac{16}{9}N_f-2 t^2+8\frac{t^4}{h^2}\right)
    +z^2-2 \frac{z^4}{h^2}
    \right\}\ln \left(\frac{\overline{\mu}^2}{M^2_W}\right)\notag
    \\
    &-\frac{8}{9}\pi i (4 N_f-3)
    \Bigg],
    \\
    \Sigma_{\rm SM}(M^2_t)&=\frac{3}{16}\frac{g^2_{0}}{(4\pi)^2}\Bigg[
    -2-4 \frac{1}{h^2} -2 h^2-\frac{256}{9}s^2 +2 t^2+16 \frac{t^4}{h^2}\notag
    \\
    &-\frac{2}{27} \left(
    39 -\frac{64}{z^2}+25 z^2 +18 \frac{z^4-1}{h^2}
    \right)\notag
    \\
    &+
    \left(
    4h^2-\frac{8}{3}t^2 +\frac{4}{3}t^2 \frac{2 t^2+h^2}{t^2-h^2}
    \right)\ln h-\frac{8}{9} \left(
    -9 \frac{z^4}{h^2}+4 \frac{(4-5z^2 +z^4)}{t^2-z^2}
    \right)\ln z\notag
    \\
    &+\left(
    \frac{128}{3}s^2 -32 \frac{t^4}{h^2}-\frac{4}{3}\frac{t^2 (2t^2+h^2)}{t^2-h^2}-\frac{32}{9}
    \frac{(z^2-1)(t^2-4)}{z^2-t^2}
    \right)\ln t\notag
    \\
    &+\frac{2}{3}\left(4t^2-h^2\right) F(M_t,M_t,M_h)
    +\frac{2}{3}\frac{(t^2+2)(t^2-1)}{t^2}F(M_t,M_W,0)\notag
    \\
    &-\frac{2}{27}\left(
    \frac{64-80z^2 +7 z^4}{z^2}+\frac{32-40 z^2 +17 z^4}{t^2}
    \right)F(M_t,M_t,M_Z)\notag
    \\
    &+\left\{
    2\left(-6 \frac{1}{h^2}-h^2-\frac{32}{3}s^2+t^2 +8 \frac{t^4}{h^2}\right)
    -\frac{4}{9}\frac{(z^2-1)(9+4h^2+9z^2)}{h^2}
    \right\}\ln \left(\frac{\overline{\mu}^2}{M^2_W}\right)
    \Bigg],
    \\
    \left(\frac{\delta g^2}{g_{0}^2}\right)_{ \rm SM}&=\frac{g_{0}^2}{(4\pi)^2} \Bigg[
    -\frac{257}{72} -\frac{1}{24}h^2 + \frac{20}{9} N_f +\frac{1}{4} t^2 -2 \ln t
    +\frac{1}{12}(12-4 h^2+h^4) F(M_W,M_h,M_W)\notag
    \\
    &-\frac{(t^2+2)(t^2-1)}{2}F(M_W,M_t,0)
    -\frac{33}{4} F(M_W,M_W,M_W)
    +\left(\frac{4}{3}N_f -\frac{43}{6}\right)\ln \left(\frac{\overline{\mu}^2}{M^2_W}\right)
    \Bigg],\label{eq:delgSM}
\end{align}
where $h:= M_h/M_W$, $t:=M_t/M_W$, $z:=M_Z/M_W$, $s:=g_s/\left(2M_W/v_0\right)$, and $N_f=3$ is the number of families.
The one-loop functions~\cite{Kajantie:1995dw} appearing in eqs.~\eqref{eq:PihSM}\text{--}\eqref{eq:delgSM} are listed as follows:
\begin{align}
    F(k,m_1,m_2)&=1-\frac{m_1^2-m_2^2}{k^2}\ln \frac{m_1}{m_2}+\frac{m_1^2+m_2^2}{m_1^2-m_2^2} \ln \frac{m_1}{m_2}\notag
    \\
    &-\frac{2}{k^2} \sqrt{(m_1+m_2)^2-k^2}\sqrt{k^2- (m_1-m_2)^2}{\rm arctan}\, \frac{\sqrt{k^2-(m_1-m_2)^2}}{(m_1+m_2)^2-k^2}
    ,
\end{align}
where its several limits are also given as~\cite{Kajantie:1995dw}
\begin{align}
    F(m_1;m_1,m_2)&=1-r^2 \frac{3-r^2}{1-r^2}\ln r -2 r\sqrt{4-r^2}{\rm arctan}\, \frac{\sqrt{2-r}}{\sqrt{2+r}},
    \\
    F(m_1;m_2,m_2)&=2-2\sqrt{4r^2-1}{\rm arctan}\, \frac{1}{\sqrt{4r^2-1}},
    \\
    F(m;m,m)&=2-\frac{\pi}{\sqrt{3}},
    \\
    F(m_1;m_2,0)&=1+(r^2-1) \ln \left(1-\frac{1}{r^2}\right),
\end{align}
with $r=m_2/m_1$.

\section{One-loop improved $\overline{\rm MS}$-parameters}
\label{sec:MSbar_para}
From eqs.~\eqref{eq:mphi}\text{--}\eqref{eq:g2}, the $\overline{{\rm MS}}$-parameters, {\it e.g.,} eqs.~\eqref{eq:m2_tree} and \eqref{eq:lambda_tree}, are improved at the one-loop level and are expressed in terms of the physical parameters as follows:
\begin{align}
m^2&=\frac{1}{2}M_h^2 +\frac{3}{4}C_{H0}v_0^4-c_{H,{\rm kin 0}}M_h^2\notag
\\
&+\frac{1}{2}{\rm Re}\, \Pi_h (M^2_h)
+12M^4_W \frac{C_{H0}}{g_0^4}\bigg[-2\frac{\delta g^2}{g_0^2}+2 \frac{{\rm Re}\, \Pi_W (M^2_W)}{M^2_W}+\frac{\delta C_H}{C_{H0}}\bigg]\notag
\\
&-c_{H,{\rm kin}0} M^2_h \bigg[\frac{{\rm Re}\, \Pi_h (M^2_h)}{M^2_h}+\frac{\delta c_{H,{\rm kin}}}{c_{H,{\rm kin}0}}\bigg],\label{eq:mu_loop}
\\
\lambda&= \frac{M_h^2}{v_0^2}+ 3 C_{H0} v_0^2-c_{H,{\rm kin}} \frac{2M_h^2}{v_0^2}
+\left[\left((C_{ll0})_{\mu ee\mu}+(C_{ll0})_{e\mu\mu e}\right)-2\left((C^{(3)}_{Hl0})_{ee}+(C^{(3)}_{Hl0})_{\mu\mu}\right)\right]\frac{M_h^2}{2}\notag
\\
&+\frac{1}{4}g_0^2 \frac{M^2_h}{M^2_W}\bigg[
    \frac{\delta g^2}{g_0^2}+\frac{{\rm Re}\, \Pi_h \left(M^2_h\right)}{M^2_h}-\frac{{\rm Re}\, \Pi_W \left(M^2_W\right)}{M^2_W}
    \bigg]\notag
    \\
    &+12M^2_W \frac{C_{H0}}{g_0^2}
    \bigg[
    -\frac{\delta g^2}{g_0^2}
    +\frac{{\rm Re}\, \Pi_W \left(M^2_W\right)}{M^2_W}+\frac{\delta C_H}{C_{H0}}
    \bigg]
    +2M^2_h C_{HW0}\bigg[\frac{{\rm Re}\, \Pi_h \left(M^2_h\right)}{M^2_h}
    +\frac{\delta C_{HW}}{C_{HW0}}
    \bigg]\notag
    \\
    &-\frac{1}{2}g^2_0 \frac{M^2_h}{M^2_W} c_{H,{\rm kin}0}\bigg[
    \frac{\delta g^2}{g_0^2}
    +\frac{{\rm Re}\, \Pi_h \left(M^2_h\right)}{M^2_h}
    -\frac{{\rm Re}\, \Pi_W \left(M^2_W\right)}{M^2_W}
    +\frac{\delta c_{H,{\rm kin}}}{c_{H,{\rm kin}0}}
    \bigg],\label{eq:lambda_loop}
    \\
    Y_t^2 & =\frac{1}{2}g_0^2 \frac{M^2_t}{M^2_W}\bigg[1+\frac{\delta g^2}{g_0^2}-\frac{{\rm Re}\, \Pi_W \left(M^2_W\right)}{M^2_W}+2 {\rm Re}\, \Sigma_t(M^2_t)\bigg]\notag
    \\
    &+4M^2_t C_{HW0}\bigg[1
    +2 {\rm Re}\, \Sigma_t(M^2_t)
    +\frac{\delta C_{HW}}{C_{HW0}}
    \bigg]\notag
    \\
    &+2\sqrt{2}M_t \frac{M_W}{g_0} C_{uH0}
    \bigg[
    1-\frac{1}{2}\frac{\delta g^2}{g_0^2}+\frac{1}{2}\frac{{\rm Re}\, \Pi_W \left(M^2_W\right)}{M^2_W}+{\rm Re}\, \Sigma_t(M^2_t)
    +\frac{\delta C_{uH}}{C_{uH0}}
    \bigg],\label{eq:Yt_loop}
\end{align}
where $c_{H,{\rm kin}0}:=\left(C_{H\Box 0}-C_{HD 0}/4\right)v^2_0$, $Y_t$ denotes top Yukawa coupling (see Appendix~\ref{sec:top_field} for its definition), and the relevant SMEFT one-loop corrections denoted with $\delta C_{\rm SMEFT}$ are listed as follows:
\begin{align}
        C_H&=C_{H0} \bigg[1+\frac{\delta C_H}{C_{H0}}\bigg],\quad c_{H,{\rm kin}} = c_{H,{\rm kin}0}\bigg[1+\frac{\delta c_{H,{\rm kin}}}{c_{H,{\rm kin}0}}\bigg],\label{eq:CHloop}
    \\
    C_{HW}&= C_{HW0}\bigg[1+\frac{\delta C_{HW}}{C_{HW0}}\bigg],\quad C_{uH}= C_{uH0}\bigg[1+\frac{\delta C_{uH}}{C_{uH0}}\bigg].\label{eq:CuHloop}
\end{align}
In the above expressions, 0 subscripts are explicitly used to denote tree-level contributions.

\section{Field-dependent masses}
\label{sec:FDM}
The following lists the SMEFT effects on the field-dependent masses of the Higgs, NG, gauge bosons (Sec.~\ref{sec:boson_field}), and top quark (Sec.~\ref{sec:top_field}).
Throughout this work, we focus on the SMEFT effects up to the first order of the dimension-six operators.
\subsection{Higgs, NG, and gauge bosons}
\label{sec:boson_field}
We first write down terms in the Lagrangian~\eqref{eq:full_L}, which are relevant to the CW contributions for the Higgs boson, the NG boson modes, as well as the $W$ and $Z$ bosons, as follows: 
\begin{align}
    \mathcal{L}_{\rm SMEFT}&\supset -\frac{1}{4}(W^I_{\mu\nu})^2-\frac{1}{4} (B_{\mu\nu})^2+(D_{\mu}H^{\dagger})(D^{\mu}H) +m^2 H^{\dagger}H-\frac{\lambda}{2} (H^{\dagger}H)^2+\mathcal{L}_{\rm gf} \notag
    \\
    &+ C_H \left(H^{\dagger}H\right)^3+C_{H\Box} (H^{\dagger}H)\Box (H^{\dagger}H) +C_{HD} (H^{\dagger}D^{\mu}H)^{\ast} (H^{\dagger}D_{\mu}H)\notag
    \\
    &
    +C_{HW} H^{\dagger} H W^I_{\mu\nu} W^{I\mu\nu}
    +C_{HB} H^{\dagger} H B_{\mu\nu}B^{\mu\nu}
    +C_{HWB} H^{\dagger} \tau^I H W^I_{\mu\nu}B^{\mu\nu},\label{eq:rel_1}
\end{align}
where $W^I_{\mu\nu}=\partial_{\mu}W_{\nu}^I-\partial_{\nu}W^I_{\mu}-g \epsilon^{IJK}W^J_{\mu}W^K_{\nu}$, and $B_{\mu\nu}=\partial_{\mu}B_{\nu}-\partial_{\nu}B_{\mu}$.
The field $H$ is an $SU(2)_L$ scalar doublet with hypercharge $Y=1/2$.
The gauge covariant derivative is defined as $D_{\mu}=\partial_{\mu} +i g_s T^A G^A_{\mu} +i g T^I W^I_{\mu}+i g' Y B_{\mu}$, where $T^A$ are the $SU(3)_c$ generators, $T^I=\tau^I/2$ are the $SU(2)_L$ generators, and $Y$ denotes the $U(1)_Y$ hypercharge.
The gauge fixing Lagrangian is defined as
\begin{align}
    \mathcal{L}_{\rm gf}=-\frac{1}{2\xi_W}(\partial^{\mu}W^I_{\mu})^2-
    \frac{1}{2\xi_B}(\partial^{\mu}B_{\mu})^2.
\end{align}

Now, let us move on to the evaluation of the field-dependent masses.
We write the complex doublet $H$ in terms of 4 real variables as follows:
\begin{align}
    H=\frac{1}{\sqrt{2}}
    \begin{pmatrix}
    \phi_1 +i\psi_1
    \\
    \phi_2 +i \psi_2
    \end{pmatrix}.
\end{align}
We here expand $H\to H+\hat{H}$ for a constant background field $\hat{H}$, where $\hat{H}=\left(0,\phi/\sqrt{2}\right)^T$.
Then, up to the second order of dynamical fields, the relevant terms of eq.~\eqref{eq:rel_1} yields the following:
\begin{align}
    \mathcal{L}_{\rm SMEFT}\supset -\frac{1}{2} \Phi_a D^{ab} \Phi_b -\frac{1}{2}V_{\mu}\Delta^{\mu\nu}V_{\nu}-\frac{1}{2}V_{\mu}M^{\mu a}\Phi_a-\frac{1}{2}\Phi_a M^{ a\mu}V_{\mu},\label{eq:sec0}
\end{align}
where the fields $\Phi_a$ and $V_{\mu}$ are defined as
\begin{align}
       \Phi_a:= 
    \left(
    \phi_1,\,\phi_2,\,\psi_1,\,\psi_2
    \right)^T,\quad V_{\mu}:= 
    \left(W^1_{\mu},\,W^2_{\mu},\,W^3_{\mu},\,B_{\mu}
    \right)^T.
\end{align}
Also, we defined the matrices as follows:
\footnotesize
\begin{align}
    D_{ab}&:={\rm diag}\Bigg(
    \Box -m^2 + \frac{\lambda}{2} \phi^2 -\frac{3}{4}C_H\phi^4,\, \left(1 -2 \left(C_{H\Box}-\frac{1}{4}C_{HD}\right)\phi^2\right)\Box -m^2 + \frac{3\lambda}{2} \phi^2 -\frac{15}{4}C_H\phi^4,\notag
    \\
    &\quad\quad\quad\quad\quad\quad\quad\quad\Box -m^2 + \frac{\lambda}{2} \phi^2-\frac{3}{4}C_H\phi^4,\,
    \left(1+\frac{1}{2}C_{HD}\phi^2\right)\Box -m^2 + \frac{\lambda}{2} \phi^2 -\frac{3}{4}C_H\phi^4
    \Bigg),
    \\
    \Delta_{\mu\nu}&:=\begin{pmatrix}
    \Delta_{\mu\nu}^{11}-\frac{g^2\phi^2}{4} g_{\mu\nu} & 0 & 0 & 0
    \\
    0 & \Delta_{\mu\nu}^{22}-\frac{g^2 \phi^2}{4} g_{\mu\nu} & 0 & 0
    \\
    0 & 0 & \Delta_{\mu\nu}^{33}-\frac{g^2 \phi^2}{4}\left(1+\frac{1}{2}\phi^2C_{HD}\right)g_{\mu\nu} & \Delta_{\mu\nu}^{3} +\frac{g' g \phi^2}{4}\left(1+\frac{1}{2}\phi^2 C_{HD}\right) g_{\mu\nu} 
    \\
    0 & 0 & \Delta_{\mu\nu}^{3}+\frac{g' g\phi^2}{4} \left(1+\frac{1}{2}\phi^2 C_{HD}\right)g_{\mu\nu} & \Delta_{\mu\nu} -\frac{{g'}^2 \phi^2}{4}\left(1+\frac{1}{2}\phi^2C_{HD}\right) g_{\mu\nu}
    \end{pmatrix}
    ,
    \\
    M_{\mu a}&:=\begin{pmatrix}
    0 & 0 & -\frac{1}{2}g \phi \partial_{\mu} & 0
    \\
    -\frac{1}{2}g \phi \partial_{\mu} & 0 & 0 & 0
    \\
    0 & 0 & 0 & \frac{1}{2}g \phi \partial_{\mu} + \frac{1}{4} \phi^3 C_{HD} g \partial_{\mu}
    \\
    0 & 0 & 0 & -\frac{1}{2}g' \phi \partial_{\mu}- \frac{1}{4}\phi^3 C_{HD} g' \partial_{\mu}
    \end{pmatrix},
\end{align}  
\normalsize
where we used the following definitions:
\begin{align}
    \Delta^{ab}_{\mu\nu} &:=- \left[\left(1-2 \phi^2 C_{HW}\right) g_{\mu\nu}\Box - \left(1-2 \phi^2 C_{HW}-\frac{1}{\xi_W}\right)\partial_{\mu}\partial_{\nu}\right]\delta^{ab},
    \\
    \Delta_{\mu\nu} &:=- \left[\left(1-2\phi^2 C_{HB}\right)g_{\mu\nu}\Box -\left(1-2\phi^2 C_{HB}-\frac{1}{\xi_B}\right)\partial_{\mu}\partial_{\nu}\right],
    \\
    \Delta_{\mu\nu}^3 &:=-\phi^2 C_{HWB} \left[g_{\mu\nu}\Box-\partial_{\mu}\partial_{\nu}\right].
\end{align}
To normalize the kinetic terms, we redefine the fields and gauge couplings as 
\begin{align}
    \phi_2 &= \bar{\phi}_2 \left[1+ \left(C_{H\Box}-\frac{1}{4}C_{HD}\right)\phi^2\right],\quad\psi_2  = \bar{\psi}_2 \left(1-\frac{1}{4}C_{HD} \phi^2\right), 
    \\
    W^I_{\mu} &= \mathcal{W}^I_{\mu} \left(1+C_{HW}\phi^2\right),~~ B_{\mu} = \mathcal{B}_{\mu} \left(1+C_{HB} \phi^2\right),~~\overline{g} = g \left(1 +C_{HW}\phi^2 \right),~~ \overline{g}' = g' \left(1+C_{HB} \phi^2\right).
\end{align}
Then, eq.~\eqref{eq:sec0} can be rewritten as
\begin{align}
    \mathcal{L}_{\rm SMEFT}\supset -\frac{1}{2}\overline{\Psi} \begin{pmatrix}
    \overline{D}_{ab} & \overline{M}_{a \mu }
    \\
    \overline{M}_{\mu a} & \overline{\Delta}_{\mu\nu}
    \end{pmatrix}
    \overline{\Psi},\label{eq:sec1}
\end{align}
where $\overline{\Psi}:= \left(\overline{\Phi}_a,~\mathcal{V}_{\mu}\right)^T$ with the following quantities:
\begin{align}
    \overline{\Phi}_a=\left(
    \phi_1,\,\overline{\phi}_2,\,\psi_1,\,\overline{\psi}_2
    \right)^T,\quad
    \mathcal{V}_{\mu}= 
    \left(
    \mathcal{W}^1_{\mu},\,\mathcal{W}^2_{\mu},\,\mathcal{W}^3_{\mu},\,\mathcal{B}_{\mu}
    \right)^T,
\end{align}
\footnotesize
\begin{align}
    \overline{D}_{ab}&:={\rm diag}\Bigg(\Box -m^2 + \frac{\lambda}{2} \phi^2-\frac{3}{4}C_H\phi^4,\,\Box +\left(-m^2 + \frac{3\lambda}{2} \phi^2\right)\left(1+2 \left(C_{H\Box}-\frac{1}{4}C_{HD}\right)\phi^2\right)-\frac{15}{4}C_H\phi^4\Bigg),\notag
    \\
    & \quad\quad\quad\quad\quad\quad\Box -m^2 + \frac{\lambda}{2} \phi^2-\frac{3}{4}C_H\phi^4,\, \Box +\left(-m^2 + \frac{\lambda}{2} \phi^2\right)\left(1-\frac{1}{2}C_{HD}\phi^2\right)-\frac{3}{4}C_H \phi^4
    \Bigg),
    \\
    \overline{M}_{\mu a}&:=\begin{pmatrix}
    0 & 0 & -\frac{1}{2}\bar{g} \phi \partial_{\mu} & 0
    \\
    -\frac{1}{2}\bar{g} \phi \partial_{\mu} & 0 & 0 & 0
    \\
    0 & 0 & 0 & \left(1-\frac{C_{HD}\phi^2}{4}\right)\left(\frac{1}{2}\bar{g} \phi \partial_{\mu} + \frac{1}{4} \phi^3 C_{HD} \bar{g} \partial_{\mu}\right)
    \\
    0 & 0 & 0 & \left(1-\frac{C_{HD}\phi^2}{4}\right)\left(-\frac{1}{2}\bar{g}' \phi \partial_{\mu}- \frac{1}{4}\phi^3 C_{HD} \bar{g}' \partial_{\mu}\right)
    \end{pmatrix},
    \\
    \overline{\Delta}_{\mu\nu}&:=\begin{pmatrix}
    \bar{\Delta}_{\mu\nu}^{11}-\frac{\bar{g}^2 \phi^2}{4} g_{\mu\nu} & 0 & 0 & 0
    \\
    0 & \bar{\Delta}_{\mu\nu}^{22}-\frac{\bar{g}^2 \phi^2}{4} g_{\mu\nu} & 0 & 0
    \\
    0 & 0 & \bar{\Delta}_{\mu\nu}^{33}-\frac{\bar{g}^2 \phi^2}{4} \left(1+\frac{\phi^2}{2}C_{HD}\right)g_{\mu\nu} & \Delta_{\mu\nu}^{3} +\frac{\bar{g}' \bar{g} \phi^2}{4}\left(1+\frac{\phi^2}{2}C_{HD}\right) g_{\mu\nu} 
    \\
    0 & 0 & \Delta_{\mu\nu}^{3}+\frac{\bar{g}' \bar{g}\phi^2}{4}\left(1+\frac{\phi^2}{2}C_{HD}\right) g_{\mu\nu} & \bar{\Delta}_{\mu\nu} -\frac{{\bar{g'}}^2 \phi^2}{4}\left(1+\frac{\phi^2}{2}C_{HD}\right) g_{\mu\nu}
    \end{pmatrix},
\end{align}
\normalsize
\begin{align}
    \overline{\Delta}^{ab}_{\mu\nu} &:=- \left[ g_{\mu\nu}\Box - \left(1-\frac{1+2 \phi^2 C_{HW}}{\xi_W}\right)\partial_{\mu}\partial_{\nu}\right]\delta^{ab},\quad\overline{\Delta}_{\mu\nu} :=- \left[g_{\mu\nu}\Box -\left(1-\frac{1+2\phi^2 C_{HB}}{\xi_B}\right)\partial_{\mu}\partial_{\nu}\right].
\end{align}
By diagonalizing eq.~\eqref{eq:sec1}, the field-dependent masses are obtained as follows: 
\begin{align}
    M_W^2(\phi)&=\frac{1}{4}\bar{g}^2 \phi^2,\label{eq:MW_phi}
    \\
    M_{Z}^2(\phi)&=\frac{1}{4}\left({\bar{g'}}^2+\bar{g}^2\right)\phi^2+\frac{1}{8}C_{HD}\left({\bar{g'}}^2+\bar{g}^2\right)\phi^4+\frac{1}{2}C_{HWB}\bar{g}'\bar{g} \phi^4,
    \\
    M_\gamma^2(\phi)&=0,
    \\
    M^2_{h}(\phi)&=-m^2+\frac{3\lambda}{2} \phi^2-\frac{15}{4}C_H \phi^4 +2C_{H\Box} \phi^2\left(-m^2 + \frac{3\lambda}{2} \phi^2\right)-\frac{1}{2}C_{HD} \phi^2 \left(-m^2 + \frac{3\lambda}{2} \phi^2\right),
    \\
    (M_1^\pm(\phi))^2&=\frac{1}{2}\bigg(m_{\chi}^2-\frac{1}{2}C_{HD}\phi^2 m^2_{\chi}\notag
    \\
    &\pm\sqrt{m_{\chi}^2\left(m_{\chi}^2-({\overline{g'}}^2\xi_B+\overline{g}^2\xi_W)\phi^2-C_{HD}\phi^2m^2_{\chi}+2C_{HB}\phi^4 {\overline{g'}}^2\xi_B+2C_{HW}\phi^4 \overline{g}^2\xi_W\right)}\bigg),
    \\
    (M_2^{\pm}(\phi))^2&=\frac{1}{2}\left(m_{\chi}^2\pm \sqrt{m_{\chi}^2 \left(m_{\chi}^2-\xi_W \overline{g}^2\phi^2 (1-2C_{HW}\phi^2)\right)}\right),\label{eq:M2_phi}
\end{align}
where $m_{\chi}^2=-m^2+\frac{\lambda}{2} \phi^2 -\frac{3}{4}C_H \phi^4$.

\subsection{Top quark}
\label{sec:top_field}
The Lagrangian including the Yukawa matrices is as follows:
\begin{align}
    \mathcal{L} &= -\left[
    H^{\dagger} \bar{d}_r [Y_d]_{rs} q_s +
    \tilde{H}^{\dagger} \bar{u}_r [Y_u]_{rs} q_s
    + H^{\dagger} \bar{e}_r [Y_e]_{rs} l_s +{\rm h.c.}
    \right]\notag
    \\
    &+\left[
    (C^{\ast}_{dH})_{sr} (H^{\dagger}H) H^{\dagger} \bar{d}_r q_s +
    (C_{uH}^{\ast})_{sr} (H^{\dagger}H) \tilde{H}^{\dagger} \bar{u}_r q_s
    +(C_{eH}^{\ast})_{sr} (H^{\dagger}H) H^{\dagger} \bar{e}_r l_s
    +{\rm h.c.}
    \right],
\end{align}
where $\tilde{H}_j =\epsilon_{jk} H^{\dagger k} $.
The fermion mass matrices are,
\begin{align}
    [M_{\psi}]_{rs} = \frac{v}{\sqrt{2}}\left([Y_{\psi}]_{rs}-
    \frac{1}{2}v^2 (C^{\ast}_{\psi H})_{sr}\right),~~~\psi=u,d,e
\end{align}
Therefore, the field-dependent mass of the top quark is given by,
\begin{align}
    M_t^2(\phi)=\frac{1}{2}\phi^2 \left(Y_t -\frac{1}{2}\phi^2 (C^{\ast}_{u H})_{33}\right)^2.\label{eq:Mt_phi}
\end{align}

\section{Thermally resummed masses}
\label{sec:resum}
The thermally resummed masses appearing in eq.~\eqref{eq:Vfull} are listed below:
\begin{align}
    &M^2_{W,{\rm res.}}(\phi,T)=M^2_W(\phi)+m_{\rm D}^2,
    \\
    &M^2_{Z,{\rm res.}}(\phi,T)= \frac{1}{8}\Bigg[
    (\bar{g'}^2+\bar{g}^2)\phi^2\left(1+\frac{C_{HD}\phi^2}{2}\right)
    +4 (m_{\rm D}^2+{m'_{\rm D}}^2)\notag
    \\
    &+4\sqrt{\left(m_{\rm D}^2+{m'}^2_{\rm D}+\frac{(\bar{g'}^2+\bar{g}^2)\phi^2}{4}\left(1+\frac{C_{HD}\phi^2}{2}\right)\right)^2-4 m_{\rm D}^2 {m'}^2_{\rm D}- (m_{\rm D}^2 \bar{g'}^2+{m'}^2_{\rm D}\bar{g}^2 )\phi^2\left(1+\frac{C_{HD}\phi^2}{2}\right)}
    \Bigg],
    \\
    &M^2_{A',{\rm res.}}(\phi,T)=
    \frac{1}{8}\Bigg[
    (\bar{g'}^2+\bar{g}^2)\phi^2\left(1+\frac{C_{HD}\phi^2}{2}\right)
    +4 (m_{\rm D}^2+{m'_{\rm D}}^2)\notag
    \\
    &-4\sqrt{\left(m_{\rm D}^2+{m'}^2_{\rm D}+\frac{(\bar{g'}^2+\bar{g}^2)\phi^2}{4}\left(1+\frac{C_{HD}\phi^2}{2}\right)\right)^2-4 m_{\rm D}^2 {m'}^2_{\rm D}- (m_{\rm D}^2 \bar{g'}^2+{m'}^2_{\rm D}\bar{g}^2 )\phi^2\left(1+\frac{C_{HD}\phi^2}{2}\right)}
    \Bigg]
    ,
    \\
    &M^2_{h,{\rm res.}}(\phi,T)=\mu^2_{{\rm res.}}+3 \lambda_{\rm res.}\phi^2-\frac{15}{4}C_H \phi^4+2 C_{H\Box}\phi^2 (\mu^2_{{\rm res.}}+3 \lambda_{\rm res.}\phi^2)
    -\frac{C_{HD} \phi^2}{2} (\mu^2_{{\rm res.}}+3 \lambda_{\rm res.}\phi^2)
    ,
    \\
    &M^2_{1\pm,{\rm res.}}(\phi,T)=\frac{1}{2}\Bigg[m_{\chi,{\rm res.}}^2-\frac{C_{HD}\phi^2}{2} m_{\chi,{\rm res.}}^2\notag
    \\
    &\pm\sqrt{m_{\chi,{\rm res.}}^2\left(m_{\chi,{\rm res.}}^2-(\bar{g'}^2\xi_B+\bar{g}^2\xi_W)\phi^2-C_{HD}\phi^2m_{\chi,{\rm res.}}^2+2C_{HB}\phi^4 \bar{g'}^2\xi_B+2C_{HW}\phi^4 \bar{g}^2\xi_W\right)}\Bigg],
    \\
    &M^2_{2\pm,{\rm res.}}(\phi,T)=\frac{1}{2}\Bigg[m_{\chi,{\rm res.}}^2\pm \sqrt{m_{\chi,{\rm res.}}^2 \left(m_{\chi,{\rm res.}}^2-\xi_W \bar{g}^2\phi^2 (1-2C_{HW}\phi^2)\right)}\Bigg],
    \\
    &m_{\chi,{\rm res.}}^2=\mu^2_{{\rm res.}}+\lambda_{\rm res.} \phi^2 -\frac{3}{4}C_H \phi^4,
    \\
    &m_{\rm D}^2 =T^2 g^2 \left(\frac{5}{6}+\frac{1}{3}N_f\right),
    \\
    &{m'}_{\rm D} ^2=T^2 {g'}^2 \left(
    \frac{1}{6} +\frac{5}{9}N_f
    \right),
    \\
    &\mu^2_{{\rm res.}}=-m^2 +\frac{T^2}{12}\left(
    3 \lambda +\frac{3}{4} \left(3 g^2 +{g'}^2\right)
    +3 Y_t^2
    \right)-\frac{1}{4} T^4 C_H,
    \\
    &\lambda_{\rm res.}=\frac{\lambda}{2} -T^2 C_H.
\end{align}
where $N_f=3$ denotes the number of kinematically active families at electroweak-scale temperatures.
Also please see Ref.~\cite{Croon:2020cgk}.

\clearpage
\bibliography{PT2.bib}

\end{document}